\documentclass[aps,prl,twocolumn,superscriptaddress,floatfix,nofootinbib]{revtex4-2}
\usepackage{amsmath,amssymb,bm,graphicx,color,booktabs}
\newcommand{\rev}[1]{#1}

\definecolor{orange}{rgb}{1,0.5,0}

\definecolor{exp}{rgb}{1,0,0}
\newcommand{\nck}[1]{#1}
\definecolor{brown}{rgb}{0.55,0.27,0.07}

\newcommand{\mg}[1]{#1}
\newcommand{\red}[1]{#1}
\newcommand{\blue}[1]{#1}
\definecolor{dkgreen}{rgb}{0,0.5,0}
\newcommand{\green}[1]{#1}
\definecolor{dkorange}{rgb}{0.9,0.45,0}
\newcommand{\orange}[1]{#1}
\definecolor{dkteal}{rgb}{0,0.5,0.5}
\definecolor{nbrown}{rgb}{0.6,0.3,0.0}

\newcommand{\gel}{\gamma^{\rm el}}
\newcommand{\Gph}{\gamma^{\rm ph}}
\newcommand{\Pconv}{P_{\rm conv}}
\newcommand{\dN}{\delta N}
\newcommand{\al}{\alpha}
\begin{document}

\title{Chirality-Induced Angular-Momentum Pumping by Moving Charges}

\author{Jun-ichiro Kishine}
\affiliation{Division of Natural and Environmental Sciences, The Open University of Japan, Chiba 261-8586, Japan}
\affiliation{Institute for Molecular Science, Okazaki, Aichi 444-8585, Japan}
\author{Ron Naaman}
\affiliation{Department of Chemical and Biological Physics, Weizmann Institute of Science, Rehovot 7610001, Israel}
\date{\today}

\begin{abstract}
Chirality-induced spin selectivity is established even in light elements at room temperature; dissipation and electron correlations are implicated, here combined in one mechanism. Injected electron wave packets coherently radiate chiral phonons; electronic absorption competing with phonon decay sets the fraction electrons receive and its temperature dependence. Spin–orbit coupling itself is not renormalized; it becomes effectively large because both the orbital and spin moments it couples are enhanced by Coulomb correlations. 
\end{abstract}

\maketitle

\textit{Introduction.}---Chirality-induced spin selectivity (CISS) is the phenomenon in which electrons passing through a chiral material acquire a spin polarization whose sign is fixed by the handedness of the structure and the direction of propagation \cite{Naaman2012,Bloom2024}.
\nck{Discovered in photoelectron transmission, with a room-temperature polarization of 60\% in DNA monolayers \cite{Ray1999,Gohler2011}, it has been extended to conduction, intramolecular electron transfer \cite{Eckvahl2023}, and nonlocal transport in chiral crystals such as CrNb$_3$S$_6$ \cite{Inui2020} and NbSi$_2$ \cite{Shiota2021}.

Four features characterize the effect: (i) the sign of the spin polarization reverses with the enantiomer and with the direction of the electron's propagation, (ii) it is robust at room temperature and may even increase with temperature, (iii) it appears in light elements with weak spin--orbit coupling (SOC), and (iv) it appears in photoelectron, conduction, and nonlocal transport alike.}
The effect is thus well established experimentally; what is missing is a single scheme that accounts for all four features.

Many mechanisms have been proposed \cite{Evers2022,Nuomin2026}\red{, but none has accounted for all observations}.
A central obstacle is a no-go theorem: time-reversal symmetry forbids spin polarization in coherent two-terminal transport \rev{for a single channel \cite{Bardarson2008}, although multi-orbital junctions can evade it} \cite{GuoSun2012,Matityahu2016,Yang2019,Yang2020,Utsumi2020}.
It has been circumvented by coupling to nuclear motion and dissipation \cite{Fransson2020,Kato2022,Volosniev2021,Vittmann2023,Das2022,Fransson2022,Upadhyay2026}, or by selecting orbital angular momentum without SOC and converting it into spin at an interface \cite{Gersten2013,Michaeli2019,Liu2021,Cho2026,Yang2025}.
These treatments include the electron--electron interaction, the electron--phonon interaction, and dissipation, either separately or in combination. Here we demonstrate that these ingredients enter as successive steps of one chain, from radiation through dissipation to amplification.

\nck{Formulating CISS as a linear response reduces it to an Edelstein-type current-induced magnetization \cite{Edelstein1990}, which requires a Fermi surface, does not grow with temperature, and is absent in gapped and molecular systems. Hence the Edelstein effect is distinct from CISS.}
\rev{The central claim of this Letter is that CISS is instead a nonequilibrium pumping process, driven by the unidirectional injection of the flow itself, of charge or of heat, and carrying a memory time.}

Dissipation, nuclear motion, and the temperature dependence all point to phonons\red{, and specifically to those that carry angular momentum} \cite{Vonsovskii1962,ZhangNiu2015,Juraschek2025}\red{. In} chiral crystals they split according to crystal angular momentum (CAM) and handedness \cite{Kishine2020,Tsunetsugu2023,KatoKishine2023,Ishito2023,Ueda2023}. Thermally driven, they \red{were shown to} inject angular momentum into an adjacent metal \cite{Hamada2020,Fransson2023,Funato2024,Yao2025,Sato2026,Yao2026,Nabei2026}. In helical crystals, the local rotation of \red{bonds transfers it} to $p$ electrons through a CAM selection rule \cite{Tateishi2025,Tateishi2026,KatoYokoshi2026}.
In these works the nonequilibrium chiral phonons are given from the outset \cite{Sun2025}.
\nck{In CISS, what is given is a directed flow, an electron current \cite{Gohler2011,Eckvahl2023,Inui2020,Shiota2021} or a heat current \cite{Ohe2024}; the questions are how such a flow excites chiral phonons with net angular momentum, and which electronic degree of freedom receives it, with what amplification.}

\begin{figure}[t]
\centering
\includegraphics[width=\columnwidth]{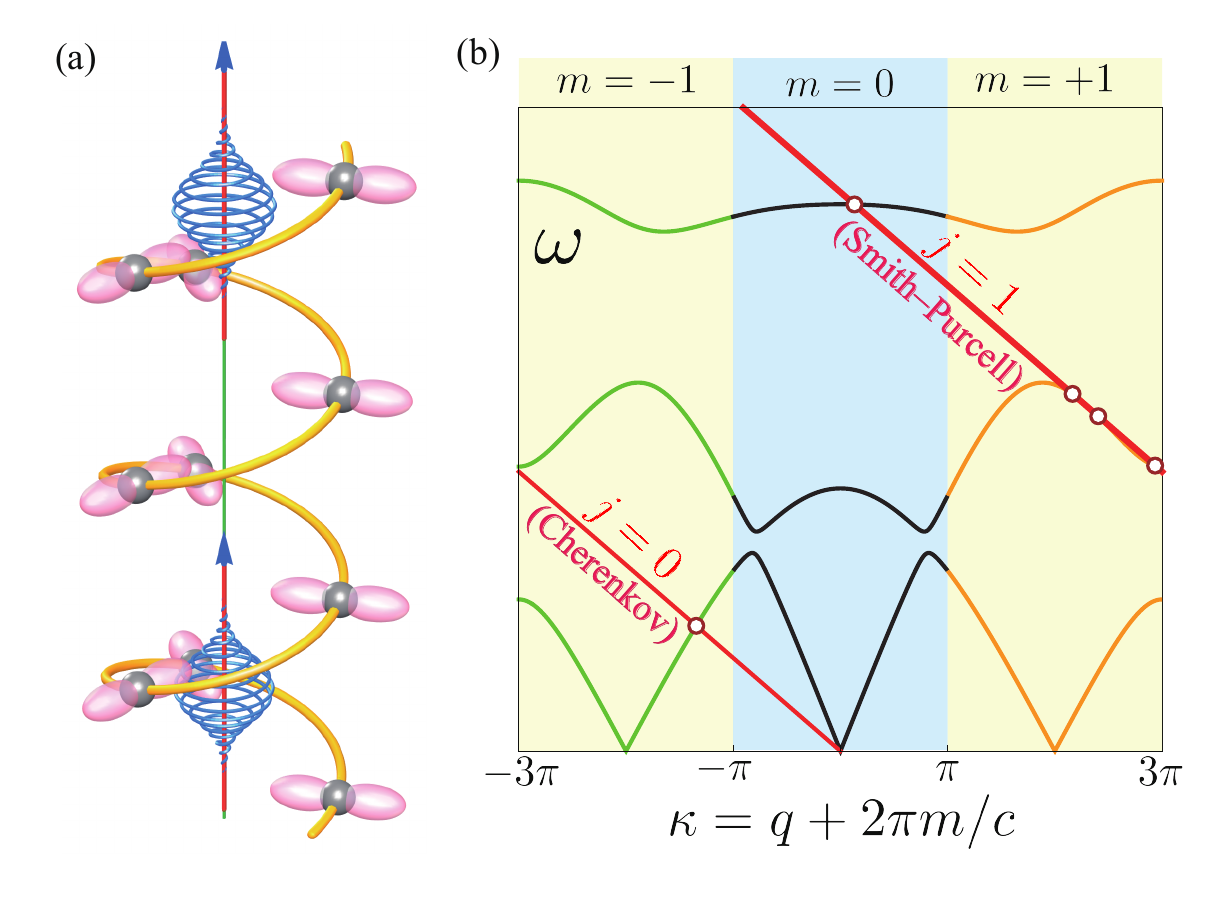}
\caption{Chirality-induced angular-momentum pumping (CAMP).
(a) A Poisson train of electron wave packets along the axis of a $3_1$ helix of $p$ orbitals.
(b) Phonon dispersion in the extended zone, $\kappa=q+2\pi m/c$, with crystal angular momentum \blue{$m=-1,0,+1$ (green, black, and orange branches; the $m=0$ sector is shaded blue); the red lines} $\omega=v(G_j-\kappa)$, $G_j=6\pi j/c$, are the phase-matching conditions of a source of velocity $v$ (Cherenkov, $j=0$; Smith--Purcell, $j=1$), and circles mark the resonances.}
\label{fig1}
\end{figure}

\begin{figure*}[t]
\centering
\includegraphics[width=1.0\textwidth]{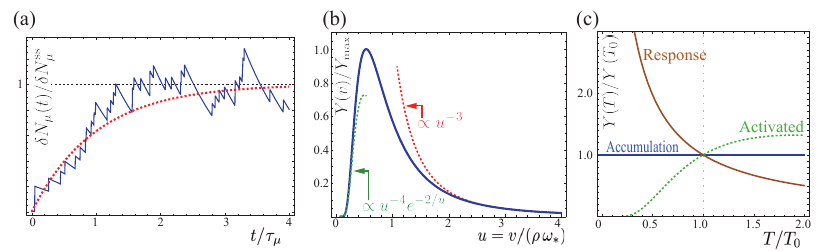}
\caption{The three factors of the yield.
(a) Occupation $\dN_\mu(t)$ of a resonant mode pumped by a Poisson train, in units of its steady-state value: one stochastic realization (blue) and the rate equation, Eq.~\eqref{eq:rate} (red); time in units of $\tau_\mu$.
(b) $Y(v)/Y_{\max}$ of the one-electron pump (blue; sector asymmetry $\propto\omega_*/v$, SM Sec.~S1.2) versus \mg{$u=v/\rho\omega_*$ ($\rho$: helix radius, $\omega_*$: frequency of the dominant resonance)}; dashed: asymptotes (SM, Sec.~S1.2).
(c) $Y(T)/Y(T_0)$ of the three types (illustrative parameters): accumulation (blue), response (brown), activated (green, dashed).}
\label{fig2}
\end{figure*}

\red{Based on this}, we develop a concrete theoretical picture, which we call chirality-induced angular-momentum pumping (CAMP), in six steps, derived in Part~I of the Supplemental Material \red{(SM)} \cite{SM} (Secs.~S1--S6, respectively).
(1) \red{The i}njected electrons are wave packets arriving at random with the Poisson statistics of shot noise, not Bloch eigenstates. Each packet exerts a classical force on the ions and radiates phase-matched coherent chiral phonons: Cherenkov radiation for the normal process and Smith--Purcell radiation for the Umklapp processes [Fig.~\ref{fig1}].
(2) Because the arrivals are random, the phase is lost and only the occupation numbers survive as a reservoir of angular momentum, whose steady state is set by the competition between the electronic absorption rate $\gel$ and the intrinsic (anharmonic) phonon decay rate $\Gph$ [Fig.~\ref{fig2}(a)].
(3) Electron--hole excitations within the Pauli window read out the reservoir through a spin-independent vertex, $\Delta L_z=\pm\hbar$ ($m\to m\pm1$).
(4) These pairs are dressed by the orbital Stoner factor $\mathcal{A}_L$.
(5) The orbital enhancement $\mathcal{A}_L$ and the triplet enhancement $\mathcal{A}_S$ share one origin, the quadrupole Slater integral $F^{(2)}$, with weights 4:1 (Hund's first and second rules); the enhancement is therefore double.
(6) A single SOC vertex connects the two, $\langle S_z\rangle=\lambda_{\rm eff}\chi^0_S\langle L_z\rangle_0$ with $\lambda_{\rm eff}=\lambda \mathcal{A}_L\mathcal{A}_S$; SOC enters only here\mg{, and it is the \red{enhancements of the orbital moment ($\mathcal{A}_L$) and of the spin response ($\mathcal{A}_S$)}, not the atomic $\lambda$, that make it appear large}.

Every transfer proceeds through occupation numbers, so the theory is a chain of rate equations.
Chirality enters twice: at step (1), where the direction of injection, a polar vector, selects the sign of the radiated angular momentum, an axial vector; and at step (3), where the rotational vertex exists only because a displacement along the helix rotates the local frame. Both are the coupling of translation and rotation characteristic of true chirality \cite{Barron2004,Kishine2022}; the enhancement and the readout do not depend on it.
The material-specific constants are left to first-principles calculations.

\textit{Model.}---We take an $n$-fold helix of $p$ shells with one atom per site, with the helical axis along $z$ [Fig.~\ref{fig1}(a)]; the screw operation $\{C_n|\tau\}$ (rotation by $2\pi/n$ about $z$, translation $\tau=pc/n$ along $z$; an $n_p$ axis with lattice constant $c$) places the $l$-th ion at $z_l=(l-1)\tau$ and azimuth $2\pi(l-1)/n$.
The Te-type $3_1$ helix ($n=3$, $p=1$) is the representative case; the disilicides ($6_2$, $6_4$) and molecular helices with large $n$, such as the $\alpha$ helix and DNA, are treated in the same framework (SM, Sec.~S1.1).
A perfect helix is the solvable reference model, with the CAM a good quantum number and exact phase matching; 
the pump itself is defined for any chiral structure with normal modes, periodic or finite, with or without screw symmetry (Appendix~A).

Electrons occupy the $p$ shell of each site ($m_\ell=0,\pm1$), the lowest shell that carries orbital angular momentum (Appendix~A). A tight-binding model on the helix \cite{Tateishi2026} gives, in order of increasing energy, the bonding, lone-pair (nonbonding), and antibonding bands.
Phonons are classified by the eigenphase $e^{-i\kappa\tau}$ of the screw operation and unfolded into the extended zone with $\kappa=q+2\pi m/(n\tau)$; $m\in\mathbb Z_n$ is the CAM, and for $3_1$, $\kappa=q+2\pi m/c$ with $m=0,\pm1$, the three sectors of Fig.~\ref{fig1}(b).
Handedness is interchanged by $p\leftrightarrow n-p$ (Appendix~A).
A mode is labeled by $\mu=(\kappa,m,\lambda)$, the axial wave vector, the CAM, and the branch, with the transverse wave vector $\bm q_\perp$ added in a three-dimensional crystal. Each mode carries a mechanical angular momentum $\ell_\mu$, and its time-reversed partner has $-\ell_\mu$.

\textit{Injected wave packets.}---An electron injected from the contact and not yet thermalized is, within its energy-relaxation length, a wave packet traveling along the axis rather than a Bloch eigenstate. 
The packet exerts on ion $l$ a screened Coulomb pulse centered at time $z_l/v$, dominated by its radial component \cite{Tateishi2025}.
Its Fourier component carries the retardation phase $e^{i\omega z_l/v}$, which interferes with the screw phase $e^{-i\kappa z_l}$ of the mode, so the lattice sum adds constructively only when $\omega_\mu=v(G_j-\kappa_\mu)$, $G_j=2\pi j/\tau$: $j=0$ is the Cherenkov condition and $j\neq0$ the Umklapp process of the Smith--Purcell type \cite{SmithPurcell1953} [Fig.~\ref{fig1}(b)] (SM, Sec.~S1.2).

On the electron side, the normal process transfers the phonon to an electron, $\bm k\to\bm k-\bm\kappa_\mu$; for small $|\kappa_\mu|$, energy conservation reads $-\hbar v_g(\bm k)\kappa_\mu=\hbar\omega_\mu=-\hbar v\kappa_\mu$, so the band velocity $v_g$ must match the velocity $v$ of the injected packet.
Since $\omega_\mu>0$, this fixes the sign of $\kappa_\mu$ for a given sign of $v$: the time-reversed partner $(-\kappa,-\ell)$ is not resonant for the same direction of incidence, and the sign of the net angular momentum is fixed at the radiation stage and reverses with the enantiomer or with the current.
The charge density of a Bloch eigenstate is stationary, so its coherent phonon amplitude vanishes, and its incoherent emission populates time-reversed partners equally and carries no net angular momentum (SM, Sec.~S1.2). Coherent radiation therefore requires the time-dependent density of a packet.

\textit{Coherent chiral phonon states.}---The force exerted by the packet couples linearly to the phonon coordinates, so one passage leaves each mode $\mu$ in a coherent state of amplitude $\al_\mu$.
The coherent occupation $|\al_\mu|^2$ is independent of temperature and fixed by the spectrum of the screened Coulomb pulse.
The angular momentum radiated per electron, $J_1=\sum_\mu\ell_\mu|\al_\mu|^2$, changes sign with the enantiomer or with $v\to-v$, and vanishes in any achiral structure, polar ones included, where a mirror plane containing the axis makes $\pm\ell$ degenerate.

The dependence on the velocity is analogous to that of the phonon-mediated attraction in superconductors. If the packet is too fast, it acts on all ions almost simultaneously and leaves no retardation. If it is too slow, the lattice follows it adiabatically. The radiation is therefore maximal in between, at $v$ of the order of the sound velocity\mg{, $v\sim\rho\omega_*$, with $\rho$ the helix radius and $\omega_*$ the frequency of the dominant resonance} [Fig.~\ref{fig2}(b)] (SM, Sec.~S1.2).
The sum is concentrated on the resonances, one for every root of $\omega_\mu=v(G_j-\kappa_\mu)$ in each sector, branch, and order $j$ [circles in Fig.~\ref{fig1}(b)]; in three dimensions each root becomes a surface over which $\bm q_\perp$ is integrated (SM, Sec.~S1.2). Roots with opposite $\ell_\mu$ partly cancel, and the net sign is that of the dominant root.

A current $I$ injects $\nu=I/e$ electrons per unit time at uncorrelated instants, a Poisson train of packets with mean spacing $e/I$ of the order of picoseconds ($\simeq1$\,ps for $I\simeq0.16\,\mu$A) [Fig.~\ref{fig1}(a)].
Since the lattice is linear, the amplitude is the superposition of the single-electron contributions with phases $e^{i\omega_\mu t_p}$, random from electron to electron because the arrival times are uncorrelated and $\omega_\mu\neq0$; the amplitudes do not add coherently over electrons, but the occupation numbers do, with mean and variance fixed by the Poisson statistics (Campbell's theorem \cite{Campbell1909}; SM, Sec.~S2.1). Thus the radiation by each electron is coherent, whereas the accumulation over electrons is incoherent.
The occupations form a reservoir that stores the excess $\dN_\mu$ over thermal equilibrium and the angular momentum $\ell_\mu\dN_\mu$; with $\dN^{(m)}\equiv\sum_{\mu\in m}\dN_\mu$ the occupation of the CAM sector $m$, time-reversal-odd observables appear only through $\dN^{(+1)}-\dN^{(-1)}$, and time-reversal-even observables in proportion to $\dN_\mu$.

\textit{Chiral phonons excite electron--hole pairs.}---A phonon rotates the bond directions at each site through the relative displacement of neighboring ions; since the hopping integrals of the $p$ orbitals are referred to the bonds, the Hamiltonian acquires, to first order, the rotational electron--phonon coupling $\propto\delta\bm\theta_l\cdot\hat{\bm L}=\hat L_-\delta\theta^{+}+\hat L_+\delta\theta^{-}$ \cite{Tateishi2026}: absorbing the circular component $\pm1$ of the displacement changes the electronic $L_z$ by $\mp\hbar$, with crystal momentum and CAM conserved.
This vertex is absent for $s$ electrons, does not depend on the spin, and is chiral: it exists because a displacement along the axis rotates the bonds, and its CAM selection rule is that of the screw.
The orbital transfer per event is set by the orbital textures of the initial and final bands, not by the mechanical $\ell_\mu$ (SM, Sec.~S3.1).

\blue{Within the $p$ shell only phonons with $m=\pm1$ (mod $n$) inject angular momentum at first order, with the net sign fixed by the chirality (Appendix~A).}

\textit{Rate equation and temperature dependence.}---The excess occupation obeys
\begin{equation}
\frac{d\,\dN_\mu}{dt}=R^{\rm pump}_\mu-\bigl(\gel_\mu+\Gph_\mu\bigr)\dN_\mu,
\label{eq:rate}
\end{equation}
where $R^{\rm pump}_\mu=\nu|\al_\mu|^2$ is fixed by the radiation alone (SM, Sec.~S2.1).
In Fig.~\ref{fig2}(a) one stochastic realization fluctuates from arrival to arrival while its mean follows Eq.~\eqref{eq:rate} and saturates at $\dN^{\rm ss}_\mu=R^{\rm pump}_\mu/(\gel_\mu+\Gph_\mu)$ within the memory time $\tau_\mu=1/(\gel_\mu+\Gph_\mu)$.
The packets accumulate in the reservoir when the memory time exceeds their spacing, $\nu\tau_\mu=I\tau_\mu/e\gtrsim1$, that is, for $I\gtrsim e/\tau_\mu$: $0.16\,\mu$A for $\tau_\mu=1$\,ps, 16\,nA for 10\,ps, and 0.16\,nA for 1\,ns ($\nu\tau_\mu=20$ in Fig.~\ref{fig2}(a)).
We stress that the essential time scale is that of the phonons. Actually, in chiral CuO the spin polarization of photoexcited carriers emerges during their scattering by phonons and reverses with the enantiomer \cite{De2026}, in direct support of the present picture.

The two rates are the two parts of the phonon self-energy, $\gel_\mu+\Gph_\mu=-2\,{\rm Im}\,\Pi_\mu(\omega_\mu)/\hbar$ (SM Fig.~S2): through $\gel_\mu$, the golden-rule absorption by an electron--hole pair (SM, Sec.~S3.1), the phonon transfers its CAM, $\Delta L_z=\pm\hbar$, to the electrons; through $\Gph_\mu$, the anharmonic decay within the lattice (Klemens decay \cite{Klemens1966}), it is dissipated.
The electrons must pick up the angular momentum before the phonon decays by itself, and the fraction they succeed in picking up is the conversion probability $\Pconv{}_{,\mu}\equiv\gel_\mu/(\gel_\mu+\Gph_\mu)$, the competition of step (2). The electron--phonon coupling enters observables only through $\Pconv$.

The temperature dependence follows from one rule: the phonon side is the loss of the resonant mode to the lattice, the anharmonic decay with a single $\hbar\omega_\mu$ plus elastic scattering by defects and boundaries, which is temperature independent and only adds a constant; the type is set by the electron side alone, constant or gapped.
With $\Gph(T)=\Gamma_0\coth(\hbar\omega_\mu/4k_BT)$,
\begin{equation}
\Pconv{}_{,\mu}(T)=\Bigl[1+\frac{\Gamma_0}{\gel(T)}\coth\left(\frac{\hbar\omega_\mu}{4k_BT}\right)\Bigr]^{-1}.
\label{eq:Pconv}
\end{equation}
The electron side, by contrast, depends on the electronic structure through the position of the Pauli window relative to $E_F$, and this alone classifies the temperature dependence into three types. If the window contains $E_F$, $\gel=\gamma_0$ is constant and the ratio $\Gamma_0/\gamma_0$ decides: the accumulation type for $\Gamma_0/\gamma_0\ll1$ ($\Pconv\to1$, a flat signal) and the response type for $\Gamma_0/\gamma_0\gg1$ ($\Pconv\simeq\gel/\Gph$, $\propto1/T$ above $\hbar\omega_\mu/4k_B$) [Fig.~\ref{fig2}(c)]. If the window lies at a distance $E_a$ from $E_F$, $\gel=\gamma_0e^{-E_a/k_BT}$ gives the activated type, rising as $e^{-E_a/k_BT}$, peaking at $k_BT\simeq E_a$, and falling as $1/T$; $E_a$ is the distance of the resonant transition from $E_F$, not a gap of the material (SM, Sec.~S7). Channels add: when a second window opens thermally, $\gel=\gamma_0[1+(\gamma_1/\gamma_0)e^{-E_a/k_BT}]$.
In every type the reservoir decays after the current is cut off with the memory time $\tau_\mu=1/(\gel+\Gph)$: the type is set by the ratio of the two rates, the memory time by their sum.

\nck{The \red{temperature-dependent }nonlocal signal of polycrystalline NbSi$_2$ \cite{Shishido2021} shows \orange{a plateau below $\simeq45$~K and a decrease above $\simeq60$~K}\rev{, \orange{which Eq.~\eqref{eq:Pconv} reproduces with a second electron channel} (Appendix~B). \blue{Because $|\al_\mu|^2$, $\langle\Delta L_z\rangle_\mu$ (single resonance), and $\eta$ (diamagnetic metal) do not depend on temperature, the measured dependence is that of $\Pconv$ alone and probes the reservoir, not the spin readout.} Where spin fluctuations are strong, as in CrNb$_3$S$_6$, $\mathcal{A}_S(T)$ adds its own dependence (Appendix~C).}}

The yield of orbital angular momentum delivered to the electrons per injected electron is
\begin{equation}
Y\equiv\frac{\dot L^{\rm inj}_z}{\nu\hbar}=\sum_\mu|\al_\mu|^2\,\Pconv{}_{,\mu}(T)\,\frac{\langle\Delta L_z\rangle_\mu}{\hbar},
\label{eq:Y}
\end{equation}
that is, the product of the number of phonons radiated per electron, the probability that a quantum goes to the electrons, and the angular momentum per quantum. The SOC does not enter $Y$.
In the accumulation limit with a single dominant resonance, the vertex, the Pauli factor, and the resonance Jacobian cancel, so $Y$ is fixed by the radiated phonon number and the orbital texture of the bands alone (SM, Sec.~S3.1).
Its velocity dependence has the optimum noted above [Fig.~\ref{fig2}(b)].

\textit{Orbital and spin ferromagnetic enhancement.}---For the $p$ shell the on-site Coulomb interaction is exactly Casimir decomposed, into $\hat N$, $\hat{\bm L}^2$, and $\hat{\bm S}^2$ (SM, Sec.~S4.1)
\begin{align}
H_{\rm int}&=\tfrac12\Bigl(F^{(0)}-\tfrac{F^{(2)}}{5}\Bigr)\hat N(\hat N-1)+\tfrac{3}{10}F^{(2)}\hat N\nonumber\\
&\quad-\tfrac{3}{50}F^{(2)}\hat{\bm L}^2-\tfrac{6}{25}F^{(2)}\hat{\bm S}^2 .
\label{eq:Hint}
\end{align}
The single quadrupole integral $F^{(2)}$ lowers the energy for large $\hat{\bm S}^2$ and large $\hat{\bm L}^2$ alike, with coefficients in the ratio 4:1: this is Hund's first and second rules in operator form, and, for comparable bare susceptibilities, it orders the two enhancements, the spin (triplet) channel before the orbital one.
In an itinerant system the Stoner factors $\mathcal{A}_{L,S}=[1-I_{L,S}\chi^0_{L,S}]^{-1}$ follow \cite{IKK1963} \blue{(Appendix~C); neither depends on the chirality.}

\textit{Readout.}---The chain from the absorbed orbital angular momentum to the spin runs through four factors, each with its own origin.
(i) The injection delivers the bare orbital angular momentum $\langle L_z\rangle_0=Y\hbar$ per packet.
(ii) The \green{static} orbital enhancement multiplies it: the enhanced orbital moment is $\mathcal{A}_L\langle L_z\rangle_0$.
(iii) A single SOC vertex $\lambda\hat{\bm L}\cdot\hat{\bm S}$ lets this moment act on the spin as an effective field $h_{\rm eff}=\lambda\mathcal{A}_L\langle L_z\rangle_0/\hbar$, which induces the bare spin $\langle S_z\rangle_0=\hbar\chi^0_Sh_{\rm eff}$, with $\chi^0_S$ the bare spin susceptibility of the spin-$\tfrac12$ electrons (units: SM, Sec.~S6.1).
(iv) The \green{static} spin enhancement multiplies it once more, so that the spin generated in the electron system per injected packet is
\begin{equation}
\mathcal{A}_S\langle S_z\rangle_0=\lambda_{\rm eff}\,\chi^0_S\,\langle L_z\rangle_0\equiv\eta\,\langle L_z\rangle_0=\eta Y\hbar,
\label{eq:eta}
\end{equation}
\mg{with
\begin{equation}
\lambda_{\rm eff}=\lambda\,\mathcal{A}_L\mathcal{A}_S .
\label{eq:lameff}
\end{equation}

This may resolve the central puzzle of CISS, namely how a weak atomic SOC can act so strongly. Since $\lambda$ is a one-body relativistic matrix element, which correlations renormalize by a factor of order unity at most, an SOC can appear larger by orders of magnitude only through the enhancement of the two moments it couples.

Equation~\eqref{eq:lameff} summarizes this statement: $\lambda_{\rm eff}$ is not a renormalized coupling but the bare $\lambda$ multiplied by the enhancement of the orbital moment it acts on, $\mathcal{A}_L$, and by the enhancement of the spin response it induces, $\mathcal{A}_S$, both of which originate from the Coulomb correlation $F^{(2)}$, can grow very large, and contain no SOC (SM, Sec.~S6.1). The giant nonlocal signal of CrNb$_3$S$_6$ \cite{Inui2020} may \red{illustrate this point}: near its helimagnetic order the spin fluctuations that build $\mathcal{A}_S$ are strong, and the same $\langle L_z\rangle_0$ is read out as a much larger spin (Appendix~C).}

The reversal of the sign with the enantiomer and with $v\to-v$ is exact by symmetry; for the dominant resonance the sign is $\mathrm{sgn}\langle S_z\rangle_0=\mathrm{sgn}(\lambda)\,\sigma_{\rm ch}\,\mathrm{sgn}(v)$, with $\sigma_{\rm ch}=\pm1$ the handedness; $\sigma_{\rm ch}$ and $\mathrm{sgn}(v)$ come from the orbital side and only $\mathrm{sgn}(\lambda)$ from SOC, and for $\lambda=0$ the enhancement $\mathcal{A}_S$ has no direction.
\mg{The SOC thus creates no angular momentum; it only converts the accumulated orbital angular momentum, enhanced by $\mathcal{A}_L$, into a spin enhanced by $\mathcal{A}_S$. The apparently large SOC in CISS is a small SOC whose effect is amplified by the Coulomb correlation.}

The spin $\eta Y\hbar$ generated per injected packet, measured in units of the spin $\hbar/2$ of one electron, defines the spin polarization, and the injection at the rate $\nu$ generates spin in the electron system at the rate
\begin{equation}
P=\frac{\eta Y\hbar}{\hbar/2}=2\eta Y,\qquad \dot S_z^{\rm gen}=\nu\,\frac{\hbar}{2}\,P.
\label{eq:P}
\end{equation}
The orbital and spin moments\green{, accumulated at the rate $\nu$ over the orbital and spin relaxation times $\tau_L$ and $\tau_s$ in the pumped volume $V$,} add to the magnetization with the $g$ factors $1$ and $2$,
\begin{equation}
M_z=-\green{\frac{\mu_B}{\hbar}\,\frac{\nu}{V}}\bigl[\tau_L\mathcal{A}_L\langle L_z\rangle_0+2\tau_s\mathcal{A}_S\langle S_z\rangle_0\bigr],
\label{eq:Mz}
\end{equation}
\nck{and the generated spin accumulates as this magnetization whether the packets form a conduction current, a photoelectron beam, or a transient displacement in a molecule.

$P$ is the yield of generated spin per injected packet; its conversion into the polarization of a transmitted current requires a readout model with escape and spin relaxation (SM, Sec.~S6.1), to be discussed elsewhere.
The readout may take place inside the medium or at an electrode with atomic SOC, which changes only $\eta$ (Appendix~C).}

\nck{\textit{Double enhancement and dichroism.}---Chirality allows a polar and an axial vector to be parallel, so a magnetization parallel to the polarization follows, in a metal from the current and in an insulating molecule from a transient charge displacement, delayed by the vibrational memory time. Both carry the double enhancement $\mathcal{A}_L\mathcal{A}_S$ of Eq.~\eqref{eq:Mz}, so the circular dichroism, which probes the orbital and spin polarizations, is expected to be amplified by the same factor (SM, Sec.~S9.2).}

\textit{Conclusion.}---We have described CISS as a chain of steps, CAMP: injected wave packets pump angular momentum into chiral phonons, the competition of electronic absorption with phonon decay sets the fraction that accumulates as orbital angular momentum, and a single SOC vertex reads it out between the two Coulomb enhancements $\mathcal{A}_L$ and $\mathcal{A}_S$. The ingredients of earlier treatments \cite{Fransson2020,Kato2022,Volosniev2021,Vittmann2023,Das2022,Fransson2022,Upadhyay2026}, electron--electron interaction, electron--vibration coupling, and dissipation, are thus steps of one process, and the SOC enhancement attributed to electron--nuclear coupling beyond the Born--Oppenheimer approximation appears as $\mathcal{A}_L\mathcal{A}_S$. Being pumping rather than linear response, the spin is not constrained by Onsager reciprocity. The peak in an $E_F$ sweep, the residual signal after the current is cut off, and the enhancement of the circular dichroism (SM Table~S1) offer tests of the proposal.
Quantitative predictions for realistic 3D materials will naturally rely on first-principles calculations of electronic and vibrational structures, for which our CAMP mechanism provides the foundational framework.

\begin{acknowledgments}
We thank Yoshihiko Togawa, Alexander S. Ovchinnikov, Tomomi Tateishi, Akihito Kato, Hiroshi M. Yamamoto, Takuya Satoh, Yusuke Kato, and Takuya Nomoto for fruitful discussions. J.~K. is especially grateful to Laurence D. Barron for first drawing his attention to this phenomenon.
J.~K. acknowledges JSPS KAKENHI Grant Nos.~JP25K00962, JP25H02149, JP23H00091, and JP26H02234, and JST ERATO Grant No.~JPMJER2503.
J.~K. also acknowledges the JSPS International Joint Research Program (JRP-LEAD with UKRI, Grant No.~JPJSJRP20241710) and the CNRS International Research Project ``Chiral Field--Matter Interaction'' (IRP ChiFiMI) for the opportunities of discussion.
Claude Fable 5.1 (Anthropic) was used for editing and checking derivations. The authors set the problem and the model, drew the conclusions, and take full responsibility for the content.
\end{acknowledgments}

\bibliographystyle{apsrev4-2}
\bibliography{CISS_refs_arXiv}

\begin{thebibliography}{66}%
\makeatletter
\providecommand \@ifxundefined [1]{%
 \@ifx{#1\undefined}
}%
\providecommand \@ifnum [1]{%
 \ifnum #1\expandafter \@firstoftwo
 \else \expandafter \@secondoftwo
 \fi
}%
\providecommand \@ifx [1]{%
 \ifx #1\expandafter \@firstoftwo
 \else \expandafter \@secondoftwo
 \fi
}%
\providecommand \natexlab [1]{#1}%
\providecommand \enquote  [1]{``#1''}%
\providecommand \bibnamefont  [1]{#1}%
\providecommand \bibfnamefont [1]{#1}%
\providecommand \citenamefont [1]{#1}%
\providecommand \href@noop [0]{\@secondoftwo}%
\providecommand \href [0]{\begingroup \@sanitize@url \@href}%
\providecommand \@href[1]{\@@startlink{#1}\@@href}%
\providecommand \@@href[1]{\endgroup#1\@@endlink}%
\providecommand \@sanitize@url [0]{\catcode `\\12\catcode `\$12\catcode
  `\&12\catcode `\#12\catcode `\^12\catcode `\_12\catcode `\%12\relax}%
\providecommand \@@startlink[1]{}%
\providecommand \@@endlink[0]{}%
\providecommand \url  [0]{\begingroup\@sanitize@url \@url }%
\providecommand \@url [1]{\endgroup\@href {#1}{\urlprefix }}%
\providecommand \urlprefix  [0]{URL }%
\providecommand \Eprint [0]{\href }%
\providecommand \doibase [0]{https://doi.org/}%
\providecommand \selectlanguage [0]{\@gobble}%
\providecommand \bibinfo  [0]{\@secondoftwo}%
\providecommand \bibfield  [0]{\@secondoftwo}%
\providecommand \translation [1]{[#1]}%
\providecommand \BibitemOpen [0]{}%
\providecommand \bibitemStop [0]{}%
\providecommand \bibitemNoStop [0]{.\EOS\space}%
\providecommand \EOS [0]{\spacefactor3000\relax}%
\providecommand \BibitemShut  [1]{\csname bibitem#1\endcsname}%
\let\auto@bib@innerbib\@empty
\bibitem [{\citenamefont {Naaman}\ and\ \citenamefont
  {Waldeck}(2012)}]{Naaman2012}%
  \BibitemOpen
  \bibfield  {author} {\bibinfo {author} {\bibfnamefont {R.}~\bibnamefont
  {Naaman}}\ and\ \bibinfo {author} {\bibfnamefont {D.~H.}\ \bibnamefont
  {Waldeck}},\ }\href@noop {} {\bibfield  {journal} {\bibinfo  {journal} {J.
  Phys. Chem. Lett.}\ }\textbf {\bibinfo {volume} {3}},\ \bibinfo {pages}
  {2178} (\bibinfo {year} {2012})}\BibitemShut {NoStop}%
\bibitem [{\citenamefont {Bloom}\ \emph {et~al.}(2024)\citenamefont {Bloom},
  \citenamefont {Paltiel}, \citenamefont {Naaman},\ and\ \citenamefont
  {Waldeck}}]{Bloom2024}%
  \BibitemOpen
  \bibfield  {author} {\bibinfo {author} {\bibfnamefont {B.~P.}\ \bibnamefont
  {Bloom}}, \bibinfo {author} {\bibfnamefont {Y.}~\bibnamefont {Paltiel}},
  \bibinfo {author} {\bibfnamefont {R.}~\bibnamefont {Naaman}},\ and\ \bibinfo
  {author} {\bibfnamefont {D.~H.}\ \bibnamefont {Waldeck}},\ }\href@noop {}
  {\bibfield  {journal} {\bibinfo  {journal} {Chem. Rev.}\ }\textbf {\bibinfo
  {volume} {124}},\ \bibinfo {pages} {1950} (\bibinfo {year}
  {2024})}\BibitemShut {NoStop}%
\bibitem [{\citenamefont {Ray}\ \emph {et~al.}(1999)\citenamefont {Ray},
  \citenamefont {Ananthavel}, \citenamefont {Waldeck},\ and\ \citenamefont
  {Naaman}}]{Ray1999}%
  \BibitemOpen
  \bibfield  {author} {\bibinfo {author} {\bibfnamefont {K.}~\bibnamefont
  {Ray}}, \bibinfo {author} {\bibfnamefont {S.~P.}\ \bibnamefont {Ananthavel}},
  \bibinfo {author} {\bibfnamefont {D.~H.}\ \bibnamefont {Waldeck}},\ and\
  \bibinfo {author} {\bibfnamefont {R.}~\bibnamefont {Naaman}},\ }\href@noop {}
  {\bibfield  {journal} {\bibinfo  {journal} {Science}\ }\textbf {\bibinfo
  {volume} {283}},\ \bibinfo {pages} {814} (\bibinfo {year}
  {1999})}\BibitemShut {NoStop}%
\bibitem [{\citenamefont {G\"ohler}\ \emph {et~al.}(2011)\citenamefont
  {G\"ohler}, \citenamefont {Hamelbeck}, \citenamefont {Markus}, \citenamefont
  {Kettner}, \citenamefont {Hanne}, \citenamefont {Vager}, \citenamefont
  {Naaman},\ and\ \citenamefont {Zacharias}}]{Gohler2011}%
  \BibitemOpen
  \bibfield  {author} {\bibinfo {author} {\bibfnamefont {B.}~\bibnamefont
  {G\"ohler}}, \bibinfo {author} {\bibfnamefont {V.}~\bibnamefont {Hamelbeck}},
  \bibinfo {author} {\bibfnamefont {T.~Z.}\ \bibnamefont {Markus}}, \bibinfo
  {author} {\bibfnamefont {M.}~\bibnamefont {Kettner}}, \bibinfo {author}
  {\bibfnamefont {G.~F.}\ \bibnamefont {Hanne}}, \bibinfo {author}
  {\bibfnamefont {Z.}~\bibnamefont {Vager}}, \bibinfo {author} {\bibfnamefont
  {R.}~\bibnamefont {Naaman}},\ and\ \bibinfo {author} {\bibfnamefont
  {H.}~\bibnamefont {Zacharias}},\ }\href@noop {} {\bibfield  {journal}
  {\bibinfo  {journal} {Science}\ }\textbf {\bibinfo {volume} {331}},\ \bibinfo
  {pages} {894} (\bibinfo {year} {2011})}\BibitemShut {NoStop}%
\bibitem [{\citenamefont {Eckvahl}\ \emph {et~al.}(2023)\citenamefont
  {Eckvahl}, \citenamefont {Tcyrulnikov}, \citenamefont {Chiesa}, \citenamefont
  {Bradley}, \citenamefont {Young}, \citenamefont {Carretta}, \citenamefont
  {Krzyaniak},\ and\ \citenamefont {Wasielewski}}]{Eckvahl2023}%
  \BibitemOpen
  \bibfield  {author} {\bibinfo {author} {\bibfnamefont {H.~J.}\ \bibnamefont
  {Eckvahl}}, \bibinfo {author} {\bibfnamefont {N.~A.}\ \bibnamefont
  {Tcyrulnikov}}, \bibinfo {author} {\bibfnamefont {A.}~\bibnamefont {Chiesa}},
  \bibinfo {author} {\bibfnamefont {J.~M.}\ \bibnamefont {Bradley}}, \bibinfo
  {author} {\bibfnamefont {R.~M.}\ \bibnamefont {Young}}, \bibinfo {author}
  {\bibfnamefont {S.}~\bibnamefont {Carretta}}, \bibinfo {author}
  {\bibfnamefont {M.~D.}\ \bibnamefont {Krzyaniak}},\ and\ \bibinfo {author}
  {\bibfnamefont {M.~R.}\ \bibnamefont {Wasielewski}},\ }\href@noop {}
  {\bibfield  {journal} {\bibinfo  {journal} {Science}\ }\textbf {\bibinfo
  {volume} {382}},\ \bibinfo {pages} {197} (\bibinfo {year}
  {2023})}\BibitemShut {NoStop}%
\bibitem [{\citenamefont {Inui}\ \emph {et~al.}(2020)\citenamefont {Inui},
  \citenamefont {Aoki}, \citenamefont {Nishiue}, \citenamefont {Shiota},
  \citenamefont {Kousaka}, \citenamefont {Shishido}, \citenamefont {Hirobe},
  \citenamefont {Suda}, \citenamefont {Ohe}, \citenamefont {Kishine},
  \citenamefont {Yamamoto},\ and\ \citenamefont {Togawa}}]{Inui2020}%
  \BibitemOpen
  \bibfield  {author} {\bibinfo {author} {\bibfnamefont {A.}~\bibnamefont
  {Inui}}, \bibinfo {author} {\bibfnamefont {R.}~\bibnamefont {Aoki}}, \bibinfo
  {author} {\bibfnamefont {Y.}~\bibnamefont {Nishiue}}, \bibinfo {author}
  {\bibfnamefont {K.}~\bibnamefont {Shiota}}, \bibinfo {author} {\bibfnamefont
  {Y.}~\bibnamefont {Kousaka}}, \bibinfo {author} {\bibfnamefont
  {H.}~\bibnamefont {Shishido}}, \bibinfo {author} {\bibfnamefont
  {D.}~\bibnamefont {Hirobe}}, \bibinfo {author} {\bibfnamefont
  {M.}~\bibnamefont {Suda}}, \bibinfo {author} {\bibfnamefont {J.}~\bibnamefont
  {Ohe}}, \bibinfo {author} {\bibfnamefont {J.}~\bibnamefont {Kishine}},
  \bibinfo {author} {\bibfnamefont {H.~M.}\ \bibnamefont {Yamamoto}},\ and\
  \bibinfo {author} {\bibfnamefont {Y.}~\bibnamefont {Togawa}},\ }\href@noop {}
  {\bibfield  {journal} {\bibinfo  {journal} {Phys. Rev. Lett.}\ }\textbf
  {\bibinfo {volume} {124}},\ \bibinfo {pages} {166602} (\bibinfo {year}
  {2020})}\BibitemShut {NoStop}%
\bibitem [{\citenamefont {Shiota}\ \emph {et~al.}(2021)\citenamefont {Shiota},
  \citenamefont {Inui}, \citenamefont {Hosaka}, \citenamefont {Amano},
  \citenamefont {\=Onuki}, \citenamefont {Hedo}, \citenamefont {Nakama},
  \citenamefont {Hirobe}, \citenamefont {Ohe}, \citenamefont {Kishine},
  \citenamefont {Yamamoto}, \citenamefont {Shishido},\ and\ \citenamefont
  {Togawa}}]{Shiota2021}%
  \BibitemOpen
  \bibfield  {author} {\bibinfo {author} {\bibfnamefont {K.}~\bibnamefont
  {Shiota}}, \bibinfo {author} {\bibfnamefont {A.}~\bibnamefont {Inui}},
  \bibinfo {author} {\bibfnamefont {Y.}~\bibnamefont {Hosaka}}, \bibinfo
  {author} {\bibfnamefont {R.}~\bibnamefont {Amano}}, \bibinfo {author}
  {\bibfnamefont {Y.}~\bibnamefont {\=Onuki}}, \bibinfo {author} {\bibfnamefont
  {M.}~\bibnamefont {Hedo}}, \bibinfo {author} {\bibfnamefont {T.}~\bibnamefont
  {Nakama}}, \bibinfo {author} {\bibfnamefont {D.}~\bibnamefont {Hirobe}},
  \bibinfo {author} {\bibfnamefont {J.}~\bibnamefont {Ohe}}, \bibinfo {author}
  {\bibfnamefont {J.}~\bibnamefont {Kishine}}, \bibinfo {author} {\bibfnamefont
  {H.~M.}\ \bibnamefont {Yamamoto}}, \bibinfo {author} {\bibfnamefont
  {H.}~\bibnamefont {Shishido}},\ and\ \bibinfo {author} {\bibfnamefont
  {Y.}~\bibnamefont {Togawa}},\ }\href@noop {} {\bibfield  {journal} {\bibinfo
  {journal} {Phys. Rev. Lett.}\ }\textbf {\bibinfo {volume} {127}},\ \bibinfo
  {pages} {126602} (\bibinfo {year} {2021})}\BibitemShut {NoStop}%
\bibitem [{\citenamefont {Evers}\ \emph {et~al.}(2022)\citenamefont {Evers}
  \emph {et~al.}}]{Evers2022}%
  \BibitemOpen
  \bibfield  {author} {\bibinfo {author} {\bibfnamefont {F.}~\bibnamefont
  {Evers}} \emph {et~al.},\ }\href@noop {} {\bibfield  {journal} {\bibinfo
  {journal} {Adv. Mater.}\ }\textbf {\bibinfo {volume} {34}},\ \bibinfo {pages}
  {2106629} (\bibinfo {year} {2022})}\BibitemShut {NoStop}%
\bibitem [{\citenamefont {Nuomin}\ \emph {et~al.}(2026)\citenamefont {Nuomin},
  \citenamefont {Charyshnikova}, \citenamefont {Song}, \citenamefont {Sun},
  \citenamefont {Nabei}, \citenamefont {Singh}, \citenamefont {Beratan} \emph
  {et~al.}}]{Nuomin2026}%
  \BibitemOpen
  \bibfield  {author} {\bibinfo {author} {\bibfnamefont {H.}~\bibnamefont
  {Nuomin}}, \bibinfo {author} {\bibfnamefont {Z.}~\bibnamefont
  {Charyshnikova}}, \bibinfo {author} {\bibfnamefont {F.-F.}\ \bibnamefont
  {Song}}, \bibinfo {author} {\bibfnamefont {R.}~\bibnamefont {Sun}}, \bibinfo
  {author} {\bibfnamefont {Y.}~\bibnamefont {Nabei}}, \bibinfo {author}
  {\bibfnamefont {N.}~\bibnamefont {Singh}}, \bibinfo {author} {\bibfnamefont
  {D.~N.}\ \bibnamefont {Beratan}}, \emph {et~al.},\ }\href@noop {} {\bibfield
  {journal} {\bibinfo  {journal} {Annu. Rev. Phys. Chem.}\ }\textbf {\bibinfo
  {volume} {77}},\ \bibinfo {pages} {513} (\bibinfo {year} {2026})}\BibitemShut
  {NoStop}%
\bibitem [{\citenamefont {Bardarson}(2008)}]{Bardarson2008}%
  \BibitemOpen
  \bibfield  {author} {\bibinfo {author} {\bibfnamefont {J.~H.}\ \bibnamefont
  {Bardarson}},\ }\href@noop {} {\bibfield  {journal} {\bibinfo  {journal} {J.
  Phys. A: Math. Theor.}\ }\textbf {\bibinfo {volume} {41}},\ \bibinfo {pages}
  {405203} (\bibinfo {year} {2008})}\BibitemShut {NoStop}%
\bibitem [{\citenamefont {Guo}\ and\ \citenamefont {Sun}(2012)}]{GuoSun2012}%
  \BibitemOpen
  \bibfield  {author} {\bibinfo {author} {\bibfnamefont {A.-M.}\ \bibnamefont
  {Guo}}\ and\ \bibinfo {author} {\bibfnamefont {Q.-F.}\ \bibnamefont {Sun}},\
  }\href@noop {} {\bibfield  {journal} {\bibinfo  {journal} {Phys. Rev. Lett.}\
  }\textbf {\bibinfo {volume} {108}},\ \bibinfo {pages} {218102} (\bibinfo
  {year} {2012})}\BibitemShut {NoStop}%
\bibitem [{\citenamefont {Matityahu}\ \emph {et~al.}(2016)\citenamefont
  {Matityahu}, \citenamefont {Utsumi}, \citenamefont {Aharony}, \citenamefont
  {Entin-Wohlman},\ and\ \citenamefont {Balseiro}}]{Matityahu2016}%
  \BibitemOpen
  \bibfield  {author} {\bibinfo {author} {\bibfnamefont {S.}~\bibnamefont
  {Matityahu}}, \bibinfo {author} {\bibfnamefont {Y.}~\bibnamefont {Utsumi}},
  \bibinfo {author} {\bibfnamefont {A.}~\bibnamefont {Aharony}}, \bibinfo
  {author} {\bibfnamefont {O.}~\bibnamefont {Entin-Wohlman}},\ and\ \bibinfo
  {author} {\bibfnamefont {C.~A.}\ \bibnamefont {Balseiro}},\ }\href@noop {}
  {\bibfield  {journal} {\bibinfo  {journal} {Phys. Rev. B}\ }\textbf {\bibinfo
  {volume} {93}},\ \bibinfo {pages} {075407} (\bibinfo {year}
  {2016})}\BibitemShut {NoStop}%
\bibitem [{\citenamefont {Yang}\ \emph {et~al.}(2019)\citenamefont {Yang},
  \citenamefont {van~der Wal},\ and\ \citenamefont {van Wees}}]{Yang2019}%
  \BibitemOpen
  \bibfield  {author} {\bibinfo {author} {\bibfnamefont {X.}~\bibnamefont
  {Yang}}, \bibinfo {author} {\bibfnamefont {C.~H.}\ \bibnamefont {van~der
  Wal}},\ and\ \bibinfo {author} {\bibfnamefont {B.~J.}\ \bibnamefont {van
  Wees}},\ }\href@noop {} {\bibfield  {journal} {\bibinfo  {journal} {Phys.
  Rev. B}\ }\textbf {\bibinfo {volume} {99}},\ \bibinfo {pages} {024418}
  (\bibinfo {year} {2019})}\BibitemShut {NoStop}%
\bibitem [{\citenamefont {Yang}\ \emph {et~al.}(2020)\citenamefont {Yang},
  \citenamefont {van~der Wal},\ and\ \citenamefont {van Wees}}]{Yang2020}%
  \BibitemOpen
  \bibfield  {author} {\bibinfo {author} {\bibfnamefont {X.}~\bibnamefont
  {Yang}}, \bibinfo {author} {\bibfnamefont {C.~H.}\ \bibnamefont {van~der
  Wal}},\ and\ \bibinfo {author} {\bibfnamefont {B.~J.}\ \bibnamefont {van
  Wees}},\ }\href@noop {} {\bibfield  {journal} {\bibinfo  {journal} {Nano
  Lett.}\ }\textbf {\bibinfo {volume} {20}},\ \bibinfo {pages} {6148} (\bibinfo
  {year} {2020})}\BibitemShut {NoStop}%
\bibitem [{\citenamefont {Utsumi}\ \emph {et~al.}(2020)\citenamefont {Utsumi},
  \citenamefont {Entin-Wohlman},\ and\ \citenamefont {Aharony}}]{Utsumi2020}%
  \BibitemOpen
  \bibfield  {author} {\bibinfo {author} {\bibfnamefont {Y.}~\bibnamefont
  {Utsumi}}, \bibinfo {author} {\bibfnamefont {O.}~\bibnamefont
  {Entin-Wohlman}},\ and\ \bibinfo {author} {\bibfnamefont {A.}~\bibnamefont
  {Aharony}},\ }\href@noop {} {\bibfield  {journal} {\bibinfo  {journal} {Phys.
  Rev. B}\ }\textbf {\bibinfo {volume} {102}},\ \bibinfo {pages} {035445}
  (\bibinfo {year} {2020})}\BibitemShut {NoStop}%
\bibitem [{\citenamefont {Fransson}(2020)}]{Fransson2020}%
  \BibitemOpen
  \bibfield  {author} {\bibinfo {author} {\bibfnamefont {J.}~\bibnamefont
  {Fransson}},\ }\href@noop {} {\bibfield  {journal} {\bibinfo  {journal}
  {Phys. Rev. B}\ }\textbf {\bibinfo {volume} {102}},\ \bibinfo {pages}
  {235416} (\bibinfo {year} {2020})}\BibitemShut {NoStop}%
\bibitem [{\citenamefont {Kato}\ \emph {et~al.}(2022)\citenamefont {Kato},
  \citenamefont {Yamamoto},\ and\ \citenamefont {Kishine}}]{Kato2022}%
  \BibitemOpen
  \bibfield  {author} {\bibinfo {author} {\bibfnamefont {A.}~\bibnamefont
  {Kato}}, \bibinfo {author} {\bibfnamefont {H.~M.}\ \bibnamefont {Yamamoto}},\
  and\ \bibinfo {author} {\bibfnamefont {J.-i.}\ \bibnamefont {Kishine}},\
  }\href@noop {} {\bibfield  {journal} {\bibinfo  {journal} {Phys. Rev. B}\
  }\textbf {\bibinfo {volume} {105}},\ \bibinfo {pages} {195117} (\bibinfo
  {year} {2022})}\BibitemShut {NoStop}%
\bibitem [{\citenamefont {Volosniev}\ \emph {et~al.}(2021)\citenamefont
  {Volosniev}, \citenamefont {Alpern}, \citenamefont {Paltiel}, \citenamefont
  {Millo}, \citenamefont {Lemeshko},\ and\ \citenamefont
  {Ghazaryan}}]{Volosniev2021}%
  \BibitemOpen
  \bibfield  {author} {\bibinfo {author} {\bibfnamefont {A.~G.}\ \bibnamefont
  {Volosniev}}, \bibinfo {author} {\bibfnamefont {H.}~\bibnamefont {Alpern}},
  \bibinfo {author} {\bibfnamefont {Y.}~\bibnamefont {Paltiel}}, \bibinfo
  {author} {\bibfnamefont {O.}~\bibnamefont {Millo}}, \bibinfo {author}
  {\bibfnamefont {M.}~\bibnamefont {Lemeshko}},\ and\ \bibinfo {author}
  {\bibfnamefont {A.}~\bibnamefont {Ghazaryan}},\ }\href@noop {} {\bibfield
  {journal} {\bibinfo  {journal} {Phys. Rev. B}\ }\textbf {\bibinfo {volume}
  {104}},\ \bibinfo {pages} {024430} (\bibinfo {year} {2021})}\BibitemShut
  {NoStop}%
\bibitem [{\citenamefont {Vittmann}\ \emph {et~al.}(2023)\citenamefont
  {Vittmann}, \citenamefont {Lim}, \citenamefont {Tamascelli}, \citenamefont
  {Huelga},\ and\ \citenamefont {Plenio}}]{Vittmann2023}%
  \BibitemOpen
  \bibfield  {author} {\bibinfo {author} {\bibfnamefont {C.}~\bibnamefont
  {Vittmann}}, \bibinfo {author} {\bibfnamefont {J.}~\bibnamefont {Lim}},
  \bibinfo {author} {\bibfnamefont {D.}~\bibnamefont {Tamascelli}}, \bibinfo
  {author} {\bibfnamefont {S.~F.}\ \bibnamefont {Huelga}},\ and\ \bibinfo
  {author} {\bibfnamefont {M.~B.}\ \bibnamefont {Plenio}},\ }\href@noop {}
  {\bibfield  {journal} {\bibinfo  {journal} {J. Phys. Chem. Lett.}\ }\textbf
  {\bibinfo {volume} {14}},\ \bibinfo {pages} {340} (\bibinfo {year}
  {2023})}\BibitemShut {NoStop}%
\bibitem [{\citenamefont {Das}\ \emph {et~al.}(2022)\citenamefont {Das},
  \citenamefont {Tassinari}, \citenamefont {Naaman},\ and\ \citenamefont
  {Fransson}}]{Das2022}%
  \BibitemOpen
  \bibfield  {author} {\bibinfo {author} {\bibfnamefont {T.~K.}\ \bibnamefont
  {Das}}, \bibinfo {author} {\bibfnamefont {F.}~\bibnamefont {Tassinari}},
  \bibinfo {author} {\bibfnamefont {R.}~\bibnamefont {Naaman}},\ and\ \bibinfo
  {author} {\bibfnamefont {J.}~\bibnamefont {Fransson}},\ }\href@noop {}
  {\bibfield  {journal} {\bibinfo  {journal} {J. Phys. Chem. C}\ }\textbf
  {\bibinfo {volume} {126}},\ \bibinfo {pages} {3257} (\bibinfo {year}
  {2022})}\BibitemShut {NoStop}%
\bibitem [{\citenamefont {Fransson}(2022)}]{Fransson2022}%
  \BibitemOpen
  \bibfield  {author} {\bibinfo {author} {\bibfnamefont {J.}~\bibnamefont
  {Fransson}},\ }\href@noop {} {\bibfield  {journal} {\bibinfo  {journal} {Isr.
  J. Chem.}\ }\textbf {\bibinfo {volume} {62}},\ \bibinfo {pages} {e202200046}
  (\bibinfo {year} {2022})}\BibitemShut {NoStop}%
\bibitem [{\citenamefont {Upadhyay}\ and\ \citenamefont
  {Levy}(2026)}]{Upadhyay2026}%
  \BibitemOpen
  \bibfield  {author} {\bibinfo {author} {\bibfnamefont {V.}~\bibnamefont
  {Upadhyay}}\ and\ \bibinfo {author} {\bibfnamefont {A.}~\bibnamefont
  {Levy}},\ }\href@noop {} {\bibfield  {journal} {\bibinfo  {journal} {J. Chem.
  Phys.}\ }\textbf {\bibinfo {volume} {164}},\ \bibinfo {pages} {164303}
  (\bibinfo {year} {2026})}\BibitemShut {NoStop}%
\bibitem [{\citenamefont {Gersten}\ \emph {et~al.}(2013)\citenamefont
  {Gersten}, \citenamefont {Kaasbjerg},\ and\ \citenamefont
  {Nitzan}}]{Gersten2013}%
  \BibitemOpen
  \bibfield  {author} {\bibinfo {author} {\bibfnamefont {J.}~\bibnamefont
  {Gersten}}, \bibinfo {author} {\bibfnamefont {K.}~\bibnamefont {Kaasbjerg}},\
  and\ \bibinfo {author} {\bibfnamefont {A.}~\bibnamefont {Nitzan}},\
  }\href@noop {} {\bibfield  {journal} {\bibinfo  {journal} {J. Chem. Phys.}\
  }\textbf {\bibinfo {volume} {139}},\ \bibinfo {pages} {114111} (\bibinfo
  {year} {2013})}\BibitemShut {NoStop}%
\bibitem [{\citenamefont {Michaeli}\ and\ \citenamefont
  {Naaman}(2019)}]{Michaeli2019}%
  \BibitemOpen
  \bibfield  {author} {\bibinfo {author} {\bibfnamefont {K.}~\bibnamefont
  {Michaeli}}\ and\ \bibinfo {author} {\bibfnamefont {R.}~\bibnamefont
  {Naaman}},\ }\href@noop {} {\bibfield  {journal} {\bibinfo  {journal} {J.
  Phys. Chem. C}\ }\textbf {\bibinfo {volume} {123}},\ \bibinfo {pages} {17043}
  (\bibinfo {year} {2019})}\BibitemShut {NoStop}%
\bibitem [{\citenamefont {Liu}\ \emph {et~al.}(2021)\citenamefont {Liu},
  \citenamefont {Xiao}, \citenamefont {Koo},\ and\ \citenamefont
  {Yan}}]{Liu2021}%
  \BibitemOpen
  \bibfield  {author} {\bibinfo {author} {\bibfnamefont {Y.}~\bibnamefont
  {Liu}}, \bibinfo {author} {\bibfnamefont {J.}~\bibnamefont {Xiao}}, \bibinfo
  {author} {\bibfnamefont {J.}~\bibnamefont {Koo}},\ and\ \bibinfo {author}
  {\bibfnamefont {B.}~\bibnamefont {Yan}},\ }\href@noop {} {\bibfield
  {journal} {\bibinfo  {journal} {Nat. Mater.}\ }\textbf {\bibinfo {volume}
  {20}},\ \bibinfo {pages} {638} (\bibinfo {year} {2021})}\BibitemShut
  {NoStop}%
\bibitem [{\citenamefont {Cho}\ \emph {et~al.}(2026)\citenamefont {Cho},
  \citenamefont {Lim},\ and\ \citenamefont {Plenio}}]{Cho2026}%
  \BibitemOpen
  \bibfield  {author} {\bibinfo {author} {\bibfnamefont {N.}~\bibnamefont
  {Cho}}, \bibinfo {author} {\bibfnamefont {J.}~\bibnamefont {Lim}},\ and\
  \bibinfo {author} {\bibfnamefont {M.~B.}\ \bibnamefont {Plenio}},\
  }\href@noop {} {\bibfield  {journal} {\bibinfo  {journal} {J. Phys. Chem.
  Lett.}\ }\textbf {\bibinfo {volume} {17}},\ \bibinfo {pages} {8535} (\bibinfo
  {year} {2026})}\BibitemShut {NoStop}%
\bibitem [{\citenamefont {Yang}\ \emph {et~al.}(2025)\citenamefont {Yang},
  \citenamefont {Li}, \citenamefont {Felser},\ and\ \citenamefont
  {Yan}}]{Yang2025}%
  \BibitemOpen
  \bibfield  {author} {\bibinfo {author} {\bibfnamefont {Q.}~\bibnamefont
  {Yang}}, \bibinfo {author} {\bibfnamefont {Y.}~\bibnamefont {Li}}, \bibinfo
  {author} {\bibfnamefont {C.}~\bibnamefont {Felser}},\ and\ \bibinfo {author}
  {\bibfnamefont {B.}~\bibnamefont {Yan}},\ }\href@noop {} {\bibfield
  {journal} {\bibinfo  {journal} {Newton}\ }\textbf {\bibinfo {volume} {1}},\
  \bibinfo {pages} {100015} (\bibinfo {year} {2025})}\BibitemShut {NoStop}%
\bibitem [{\citenamefont {Edelstein}(1990)}]{Edelstein1990}%
  \BibitemOpen
  \bibfield  {author} {\bibinfo {author} {\bibfnamefont {V.~M.}\ \bibnamefont
  {Edelstein}},\ }\href@noop {} {\bibfield  {journal} {\bibinfo  {journal}
  {Solid State Commun.}\ }\textbf {\bibinfo {volume} {73}},\ \bibinfo {pages}
  {233} (\bibinfo {year} {1990})}\BibitemShut {NoStop}%
\bibitem [{\citenamefont {Vonsovskii}\ and\ \citenamefont
  {Svirskii}(1962)}]{Vonsovskii1962}%
  \BibitemOpen
  \bibfield  {author} {\bibinfo {author} {\bibfnamefont {S.~V.}\ \bibnamefont
  {Vonsovskii}}\ and\ \bibinfo {author} {\bibfnamefont {M.~S.}\ \bibnamefont
  {Svirskii}},\ }\href@noop {} {\bibfield  {journal} {\bibinfo  {journal} {Sov.
  Phys. Solid State}\ }\textbf {\bibinfo {volume} {3}},\ \bibinfo {pages}
  {1568} (\bibinfo {year} {1962})}\BibitemShut {NoStop}%
\bibitem [{\citenamefont {Zhang}\ and\ \citenamefont
  {Niu}(2015)}]{ZhangNiu2015}%
  \BibitemOpen
  \bibfield  {author} {\bibinfo {author} {\bibfnamefont {L.}~\bibnamefont
  {Zhang}}\ and\ \bibinfo {author} {\bibfnamefont {Q.}~\bibnamefont {Niu}},\
  }\href@noop {} {\bibfield  {journal} {\bibinfo  {journal} {Phys. Rev. Lett.}\
  }\textbf {\bibinfo {volume} {115}},\ \bibinfo {pages} {115502} (\bibinfo
  {year} {2015})}\BibitemShut {NoStop}%
\bibitem [{\citenamefont {Juraschek}\ \emph {et~al.}(2025)\citenamefont
  {Juraschek} \emph {et~al.}}]{Juraschek2025}%
  \BibitemOpen
  \bibfield  {author} {\bibinfo {author} {\bibfnamefont {D.~M.}\ \bibnamefont
  {Juraschek}} \emph {et~al.},\ }\href@noop {} {\bibfield  {journal} {\bibinfo
  {journal} {Nat. Phys.}\ }\textbf {\bibinfo {volume} {21}},\ \bibinfo {pages}
  {1532} (\bibinfo {year} {2025})}\BibitemShut {NoStop}%
\bibitem [{\citenamefont {Kishine}\ \emph {et~al.}(2020)\citenamefont
  {Kishine}, \citenamefont {Ovchinnikov},\ and\ \citenamefont
  {Tereshchenko}}]{Kishine2020}%
  \BibitemOpen
  \bibfield  {author} {\bibinfo {author} {\bibfnamefont {J.-i.}\ \bibnamefont
  {Kishine}}, \bibinfo {author} {\bibfnamefont {A.~S.}\ \bibnamefont
  {Ovchinnikov}},\ and\ \bibinfo {author} {\bibfnamefont {A.~A.}\ \bibnamefont
  {Tereshchenko}},\ }\href@noop {} {\bibfield  {journal} {\bibinfo  {journal}
  {Phys. Rev. Lett.}\ }\textbf {\bibinfo {volume} {125}},\ \bibinfo {pages}
  {245302} (\bibinfo {year} {2020})}\BibitemShut {NoStop}%
\bibitem [{\citenamefont {Tsunetsugu}\ and\ \citenamefont
  {Kusunose}(2023)}]{Tsunetsugu2023}%
  \BibitemOpen
  \bibfield  {author} {\bibinfo {author} {\bibfnamefont {H.}~\bibnamefont
  {Tsunetsugu}}\ and\ \bibinfo {author} {\bibfnamefont {H.}~\bibnamefont
  {Kusunose}},\ }\href@noop {} {\bibfield  {journal} {\bibinfo  {journal} {J.
  Phys. Soc. Jpn.}\ }\textbf {\bibinfo {volume} {92}},\ \bibinfo {pages}
  {023601} (\bibinfo {year} {2023})}\BibitemShut {NoStop}%
\bibitem [{\citenamefont {Kato}\ and\ \citenamefont
  {Kishine}(2023)}]{KatoKishine2023}%
  \BibitemOpen
  \bibfield  {author} {\bibinfo {author} {\bibfnamefont {A.}~\bibnamefont
  {Kato}}\ and\ \bibinfo {author} {\bibfnamefont {J.-i.}\ \bibnamefont
  {Kishine}},\ }\href@noop {} {\bibfield  {journal} {\bibinfo  {journal} {J.
  Phys. Soc. Jpn.}\ }\textbf {\bibinfo {volume} {92}},\ \bibinfo {pages}
  {075002} (\bibinfo {year} {2023})}\BibitemShut {NoStop}%
\bibitem [{\citenamefont {Ishito}\ \emph {et~al.}(2023)\citenamefont {Ishito},
  \citenamefont {Mao}, \citenamefont {Kousaka}, \citenamefont {Togawa},
  \citenamefont {Iwasaki}, \citenamefont {Zhang}, \citenamefont {Murakami},
  \citenamefont {Kishine},\ and\ \citenamefont {Satoh}}]{Ishito2023}%
  \BibitemOpen
  \bibfield  {author} {\bibinfo {author} {\bibfnamefont {K.}~\bibnamefont
  {Ishito}}, \bibinfo {author} {\bibfnamefont {H.}~\bibnamefont {Mao}},
  \bibinfo {author} {\bibfnamefont {Y.}~\bibnamefont {Kousaka}}, \bibinfo
  {author} {\bibfnamefont {Y.}~\bibnamefont {Togawa}}, \bibinfo {author}
  {\bibfnamefont {S.}~\bibnamefont {Iwasaki}}, \bibinfo {author} {\bibfnamefont
  {T.}~\bibnamefont {Zhang}}, \bibinfo {author} {\bibfnamefont
  {S.}~\bibnamefont {Murakami}}, \bibinfo {author} {\bibfnamefont {J.-i.}\
  \bibnamefont {Kishine}},\ and\ \bibinfo {author} {\bibfnamefont
  {T.}~\bibnamefont {Satoh}},\ }\href@noop {} {\bibfield  {journal} {\bibinfo
  {journal} {Nat. Phys.}\ }\textbf {\bibinfo {volume} {19}},\ \bibinfo {pages}
  {35} (\bibinfo {year} {2023})}\BibitemShut {NoStop}%
\bibitem [{\citenamefont {Ueda}\ \emph {et~al.}(2023)\citenamefont {Ueda},
  \citenamefont {Garc\'ia-Fern\'andez}, \citenamefont {Agrestini},
  \citenamefont {Romao}, \citenamefont {van~den Brink}, \citenamefont
  {Spaldin}, \citenamefont {Zhou},\ and\ \citenamefont {Staub}}]{Ueda2023}%
  \BibitemOpen
  \bibfield  {author} {\bibinfo {author} {\bibfnamefont {H.}~\bibnamefont
  {Ueda}}, \bibinfo {author} {\bibfnamefont {M.}~\bibnamefont
  {Garc\'ia-Fern\'andez}}, \bibinfo {author} {\bibfnamefont {S.}~\bibnamefont
  {Agrestini}}, \bibinfo {author} {\bibfnamefont {C.~P.}\ \bibnamefont
  {Romao}}, \bibinfo {author} {\bibfnamefont {J.}~\bibnamefont {van~den
  Brink}}, \bibinfo {author} {\bibfnamefont {N.~A.}\ \bibnamefont {Spaldin}},
  \bibinfo {author} {\bibfnamefont {K.-J.}\ \bibnamefont {Zhou}},\ and\
  \bibinfo {author} {\bibfnamefont {U.}~\bibnamefont {Staub}},\ }\href@noop {}
  {\bibfield  {journal} {\bibinfo  {journal} {Nature}\ }\textbf {\bibinfo
  {volume} {618}},\ \bibinfo {pages} {946} (\bibinfo {year}
  {2023})}\BibitemShut {NoStop}%
\bibitem [{\citenamefont {Hamada}\ and\ \citenamefont
  {Murakami}(2020)}]{Hamada2020}%
  \BibitemOpen
  \bibfield  {author} {\bibinfo {author} {\bibfnamefont {M.}~\bibnamefont
  {Hamada}}\ and\ \bibinfo {author} {\bibfnamefont {S.}~\bibnamefont
  {Murakami}},\ }\href@noop {} {\bibfield  {journal} {\bibinfo  {journal}
  {Phys. Rev. Research}\ }\textbf {\bibinfo {volume} {2}},\ \bibinfo {pages}
  {023275} (\bibinfo {year} {2020})}\BibitemShut {NoStop}%
\bibitem [{\citenamefont {Fransson}(2023)}]{Fransson2023}%
  \BibitemOpen
  \bibfield  {author} {\bibinfo {author} {\bibfnamefont {J.}~\bibnamefont
  {Fransson}},\ }\href@noop {} {\bibfield  {journal} {\bibinfo  {journal}
  {Phys. Rev. Research}\ }\textbf {\bibinfo {volume} {5}},\ \bibinfo {pages}
  {L022039} (\bibinfo {year} {2023})}\BibitemShut {NoStop}%
\bibitem [{\citenamefont {Funato}\ \emph {et~al.}(2024)\citenamefont {Funato},
  \citenamefont {Matsuo},\ and\ \citenamefont {Kato}}]{Funato2024}%
  \BibitemOpen
  \bibfield  {author} {\bibinfo {author} {\bibfnamefont {T.}~\bibnamefont
  {Funato}}, \bibinfo {author} {\bibfnamefont {M.}~\bibnamefont {Matsuo}},\
  and\ \bibinfo {author} {\bibfnamefont {T.}~\bibnamefont {Kato}},\ }\href@noop
  {} {\bibfield  {journal} {\bibinfo  {journal} {Phys. Rev. Lett.}\ }\textbf
  {\bibinfo {volume} {132}},\ \bibinfo {pages} {236201} (\bibinfo {year}
  {2024})}\BibitemShut {NoStop}%
\bibitem [{\citenamefont {Yao}\ and\ \citenamefont {Murakami}(2025)}]{Yao2025}%
  \BibitemOpen
  \bibfield  {author} {\bibinfo {author} {\bibfnamefont {D.}~\bibnamefont
  {Yao}}\ and\ \bibinfo {author} {\bibfnamefont {S.}~\bibnamefont {Murakami}},\
  }\href@noop {} {\bibfield  {journal} {\bibinfo  {journal} {Phys. Rev. B}\
  }\textbf {\bibinfo {volume} {111}},\ \bibinfo {pages} {134414} (\bibinfo
  {year} {2025})}\BibitemShut {NoStop}%
\bibitem [{\citenamefont {Sato}\ \emph {et~al.}(2026)\citenamefont {Sato},
  \citenamefont {Kato},\ and\ \citenamefont {Manchon}}]{Sato2026}%
  \BibitemOpen
  \bibfield  {author} {\bibinfo {author} {\bibfnamefont {T.}~\bibnamefont
  {Sato}}, \bibinfo {author} {\bibfnamefont {T.}~\bibnamefont {Kato}},\ and\
  \bibinfo {author} {\bibfnamefont {A.}~\bibnamefont {Manchon}},\ }\href@noop
  {} {\bibfield  {journal} {\bibinfo  {journal} {Phys. Rev. Lett.}\ }\textbf
  {\bibinfo {volume} {137}},\ \bibinfo {pages} {066302} (\bibinfo {year}
  {2026})}\BibitemShut {NoStop}%
\bibitem [{\citenamefont {Yao}\ \emph {et~al.}(2026)\citenamefont {Yao},
  \citenamefont {Go}, \citenamefont {Mokrousov},\ and\ \citenamefont
  {Murakami}}]{Yao2026}%
  \BibitemOpen
  \bibfield  {author} {\bibinfo {author} {\bibfnamefont {D.}~\bibnamefont
  {Yao}}, \bibinfo {author} {\bibfnamefont {D.}~\bibnamefont {Go}}, \bibinfo
  {author} {\bibfnamefont {Y.}~\bibnamefont {Mokrousov}},\ and\ \bibinfo
  {author} {\bibfnamefont {S.}~\bibnamefont {Murakami}},\ }\href@noop {}
  {\bibfield  {journal} {\bibinfo  {journal} {Phys. Rev. Lett.}\ }\textbf
  {\bibinfo {volume} {137}},\ \bibinfo {pages} {066301} (\bibinfo {year}
  {2026})}\BibitemShut {NoStop}%
\bibitem [{\citenamefont {Nabei}\ \emph {et~al.}(2026)\citenamefont {Nabei}
  \emph {et~al.}}]{Nabei2026}%
  \BibitemOpen
  \bibfield  {author} {\bibinfo {author} {\bibfnamefont {Y.}~\bibnamefont
  {Nabei}} \emph {et~al.},\ }\href@noop {} {\bibfield  {journal} {\bibinfo
  {journal} {Nat. Phys.}\ }\textbf {\bibinfo {volume} {22}},\ \bibinfo {pages}
  {245} (\bibinfo {year} {2026})}\BibitemShut {NoStop}%
\bibitem [{\citenamefont {Tateishi}\ \emph {et~al.}(2025)\citenamefont
  {Tateishi}, \citenamefont {Kato},\ and\ \citenamefont
  {Kishine}}]{Tateishi2025}%
  \BibitemOpen
  \bibfield  {author} {\bibinfo {author} {\bibfnamefont {T.}~\bibnamefont
  {Tateishi}}, \bibinfo {author} {\bibfnamefont {A.}~\bibnamefont {Kato}},\
  and\ \bibinfo {author} {\bibfnamefont {J.-i.}\ \bibnamefont {Kishine}},\
  }\href@noop {} {\bibfield  {journal} {\bibinfo  {journal} {J. Phys. Soc.
  Jpn.}\ }\textbf {\bibinfo {volume} {94}},\ \bibinfo {pages} {053601}
  (\bibinfo {year} {2025})}\BibitemShut {NoStop}%
\bibitem [{\citenamefont {Tateishi}\ \emph {et~al.}(2026)\citenamefont
  {Tateishi}, \citenamefont {Kato}, \citenamefont {Ovchinnikov},\ and\
  \citenamefont {Kishine}}]{Tateishi2026}%
  \BibitemOpen
  \bibfield  {author} {\bibinfo {author} {\bibfnamefont {T.}~\bibnamefont
  {Tateishi}}, \bibinfo {author} {\bibfnamefont {A.}~\bibnamefont {Kato}},
  \bibinfo {author} {\bibfnamefont {A.~S.}\ \bibnamefont {Ovchinnikov}},\ and\
  \bibinfo {author} {\bibfnamefont {J.-i.}\ \bibnamefont {Kishine}},\
  }\href@noop {} {\bibfield  {journal} {\bibinfo  {journal} {J. Phys. Soc.
  Jpn.}\ }\textbf {\bibinfo {volume} {95}},\ \bibinfo {pages} {063705}
  (\bibinfo {year} {2026})}\BibitemShut {NoStop}%
\bibitem [{\citenamefont {Kato}\ \emph {et~al.}(2026)\citenamefont {Kato},
  \citenamefont {Yokoshi},\ and\ \citenamefont {Kishine}}]{KatoYokoshi2026}%
  \BibitemOpen
  \bibfield  {author} {\bibinfo {author} {\bibfnamefont {A.}~\bibnamefont
  {Kato}}, \bibinfo {author} {\bibfnamefont {N.}~\bibnamefont {Yokoshi}},\ and\
  \bibinfo {author} {\bibfnamefont {J.-i.}\ \bibnamefont {Kishine}},\ }\href
  {https://doi.org/10.1103/kr19-7s6q} {\bibfield  {journal} {\bibinfo
  {journal} {Phys. Rev. B}\ }\textbf {\bibinfo {volume} {114}},\ \bibinfo
  {pages} {165118} (\bibinfo {year} {2026})}\BibitemShut {NoStop}%
\bibitem [{\citenamefont {Sun}\ \emph {et~al.}(2025)\citenamefont {Sun},
  \citenamefont {Chen}, \citenamefont {Wang}, \citenamefont {Zeng},
  \citenamefont {Zhou},\ and\ \citenamefont {Zhang}}]{Sun2025}%
  \BibitemOpen
  \bibfield  {author} {\bibinfo {author} {\bibfnamefont {H.}~\bibnamefont
  {Sun}}, \bibinfo {author} {\bibfnamefont {J.}~\bibnamefont {Chen}}, \bibinfo
  {author} {\bibfnamefont {Q.}~\bibnamefont {Wang}}, \bibinfo {author}
  {\bibfnamefont {X.}~\bibnamefont {Zeng}}, \bibinfo {author} {\bibfnamefont
  {J.}~\bibnamefont {Zhou}},\ and\ \bibinfo {author} {\bibfnamefont
  {L.}~\bibnamefont {Zhang}},\ }\href@noop {} {\bibfield  {journal} {\bibinfo
  {journal} {Chin. Phys. Lett.}\ }\textbf {\bibinfo {volume} {42}},\ \bibinfo
  {pages} {090711} (\bibinfo {year} {2025})}\BibitemShut {NoStop}%
\bibitem [{\citenamefont {Ohe}\ \emph {et~al.}(2024)\citenamefont {Ohe},
  \citenamefont {Shishido}, \citenamefont {Kato}, \citenamefont {Utsumi},
  \citenamefont {Matsuura},\ and\ \citenamefont {Togawa}}]{Ohe2024}%
  \BibitemOpen
  \bibfield  {author} {\bibinfo {author} {\bibfnamefont {K.}~\bibnamefont
  {Ohe}}, \bibinfo {author} {\bibfnamefont {H.}~\bibnamefont {Shishido}},
  \bibinfo {author} {\bibfnamefont {M.}~\bibnamefont {Kato}}, \bibinfo {author}
  {\bibfnamefont {S.}~\bibnamefont {Utsumi}}, \bibinfo {author} {\bibfnamefont
  {H.}~\bibnamefont {Matsuura}},\ and\ \bibinfo {author} {\bibfnamefont
  {Y.}~\bibnamefont {Togawa}},\ }\href@noop {} {\bibfield  {journal} {\bibinfo
  {journal} {Phys. Rev. Lett.}\ }\textbf {\bibinfo {volume} {132}},\ \bibinfo
  {pages} {056302} (\bibinfo {year} {2024})}\BibitemShut {NoStop}%
\bibitem [{SM()}]{SM}%
  \BibitemOpen
  \href@noop {} {}\bibinfo {note} {See Supplemental Material, appended at the
  end of this preprint, for the derivations of steps (1)--(6), the temperature
  dependence and the NbSi$_2$ analysis, and the extensions.}\BibitemShut
  {Stop}%
\bibitem [{\citenamefont {Barron}(2004)}]{Barron2004}%
  \BibitemOpen
  \bibfield  {author} {\bibinfo {author} {\bibfnamefont {L.~D.}\ \bibnamefont
  {Barron}},\ }\href@noop {} {\emph {\bibinfo {title} {Molecular Light
  Scattering and Optical Activity}}},\ \bibinfo {edition} {2nd}\ ed.\ (\bibinfo
   {publisher} {Cambridge University Press},\ \bibinfo {address} {Cambridge},\
  \bibinfo {year} {2004})\BibitemShut {NoStop}%
\bibitem [{\citenamefont {Kishine}\ \emph {et~al.}(2022)\citenamefont
  {Kishine}, \citenamefont {Kusunose},\ and\ \citenamefont
  {Yamamoto}}]{Kishine2022}%
  \BibitemOpen
  \bibfield  {author} {\bibinfo {author} {\bibfnamefont {J.-i.}\ \bibnamefont
  {Kishine}}, \bibinfo {author} {\bibfnamefont {H.}~\bibnamefont {Kusunose}},\
  and\ \bibinfo {author} {\bibfnamefont {H.~M.}\ \bibnamefont {Yamamoto}},\
  }\href@noop {} {\bibfield  {journal} {\bibinfo  {journal} {Isr. J. Chem.}\
  }\textbf {\bibinfo {volume} {62}},\ \bibinfo {pages} {e202200049} (\bibinfo
  {year} {2022})}\BibitemShut {NoStop}%
\bibitem [{\citenamefont {Smith}\ and\ \citenamefont
  {Purcell}(1953)}]{SmithPurcell1953}%
  \BibitemOpen
  \bibfield  {author} {\bibinfo {author} {\bibfnamefont {S.~J.}\ \bibnamefont
  {Smith}}\ and\ \bibinfo {author} {\bibfnamefont {E.~M.}\ \bibnamefont
  {Purcell}},\ }\href@noop {} {\bibfield  {journal} {\bibinfo  {journal} {Phys.
  Rev.}\ }\textbf {\bibinfo {volume} {92}},\ \bibinfo {pages} {1069} (\bibinfo
  {year} {1953})}\BibitemShut {NoStop}%
\bibitem [{\citenamefont {Campbell}(1909)}]{Campbell1909}%
  \BibitemOpen
  \bibfield  {author} {\bibinfo {author} {\bibfnamefont {N.~R.}\ \bibnamefont
  {Campbell}},\ }\href@noop {} {\bibfield  {journal} {\bibinfo  {journal}
  {Proc. Cambridge Philos. Soc.}\ }\textbf {\bibinfo {volume} {15}},\ \bibinfo
  {pages} {117} (\bibinfo {year} {1909})}\BibitemShut {NoStop}%
\bibitem [{\citenamefont {De}\ \emph {et~al.}(2026)\citenamefont {De},
  \citenamefont {Partin}, \citenamefont {Davis}, \citenamefont {Dykstra},
  \citenamefont {Bloom}, \citenamefont {Shiby}, \citenamefont {Waldeck},
  \citenamefont {Schultze},\ and\ \citenamefont {Baker}}]{De2026}%
  \BibitemOpen
  \bibfield  {author} {\bibinfo {author} {\bibfnamefont {R.}~\bibnamefont
  {De}}, \bibinfo {author} {\bibfnamefont {A.}~\bibnamefont {Partin}}, \bibinfo
  {author} {\bibfnamefont {C.}~\bibnamefont {Davis}}, \bibinfo {author}
  {\bibfnamefont {C.}~\bibnamefont {Dykstra}}, \bibinfo {author} {\bibfnamefont
  {B.~P.}\ \bibnamefont {Bloom}}, \bibinfo {author} {\bibfnamefont
  {E.}~\bibnamefont {Shiby}}, \bibinfo {author} {\bibfnamefont {D.~H.}\
  \bibnamefont {Waldeck}}, \bibinfo {author} {\bibfnamefont {M.}~\bibnamefont
  {Schultze}},\ and\ \bibinfo {author} {\bibfnamefont {L.~R.}\ \bibnamefont
  {Baker}},\ }\href@noop {} {\bibfield  {journal} {\bibinfo  {journal} {J. Am.
  Chem. Soc.}\ }\textbf {\bibinfo {volume} {148}},\ \bibinfo {pages} {36314}
  (\bibinfo {year} {2026})}\BibitemShut {NoStop}%
\bibitem [{\citenamefont {Klemens}(1966)}]{Klemens1966}%
  \BibitemOpen
  \bibfield  {author} {\bibinfo {author} {\bibfnamefont {P.~G.}\ \bibnamefont
  {Klemens}},\ }\href@noop {} {\bibfield  {journal} {\bibinfo  {journal} {Phys.
  Rev.}\ }\textbf {\bibinfo {volume} {148}},\ \bibinfo {pages} {845} (\bibinfo
  {year} {1966})}\BibitemShut {NoStop}%
\bibitem [{\citenamefont {Shishido}\ \emph {et~al.}(2021)\citenamefont
  {Shishido}, \citenamefont {Sakai}, \citenamefont {Hosaka},\ and\
  \citenamefont {Togawa}}]{Shishido2021}%
  \BibitemOpen
  \bibfield  {author} {\bibinfo {author} {\bibfnamefont {H.}~\bibnamefont
  {Shishido}}, \bibinfo {author} {\bibfnamefont {R.}~\bibnamefont {Sakai}},
  \bibinfo {author} {\bibfnamefont {Y.}~\bibnamefont {Hosaka}},\ and\ \bibinfo
  {author} {\bibfnamefont {Y.}~\bibnamefont {Togawa}},\ }\href@noop {}
  {\bibfield  {journal} {\bibinfo  {journal} {Appl. Phys. Lett.}\ }\textbf
  {\bibinfo {volume} {119}},\ \bibinfo {pages} {182403} (\bibinfo {year}
  {2021})}\BibitemShut {NoStop}%
\bibitem [{\citenamefont {Izuyama}\ \emph {et~al.}(1963)\citenamefont
  {Izuyama}, \citenamefont {Kim},\ and\ \citenamefont {Kubo}}]{IKK1963}%
  \BibitemOpen
  \bibfield  {author} {\bibinfo {author} {\bibfnamefont {T.}~\bibnamefont
  {Izuyama}}, \bibinfo {author} {\bibfnamefont {D.-J.}\ \bibnamefont {Kim}},\
  and\ \bibinfo {author} {\bibfnamefont {R.}~\bibnamefont {Kubo}},\ }\href@noop
  {} {\bibfield  {journal} {\bibinfo  {journal} {J. Phys. Soc. Jpn.}\ }\textbf
  {\bibinfo {volume} {18}},\ \bibinfo {pages} {1025} (\bibinfo {year}
  {1963})}\BibitemShut {NoStop}%
\bibitem [{\citenamefont {D'yakonov}\ and\ \citenamefont
  {Perel'}(1971)}]{Dyakonov1971}%
  \BibitemOpen
  \bibfield  {author} {\bibinfo {author} {\bibfnamefont {M.~I.}\ \bibnamefont
  {D'yakonov}}\ and\ \bibinfo {author} {\bibfnamefont {V.~I.}\ \bibnamefont
  {Perel'}},\ }\href@noop {} {\bibfield  {journal} {\bibinfo  {journal} {Zh.
  Eksp. Teor. Fiz.}\ }\textbf {\bibinfo {volume} {60}},\ \bibinfo {pages}
  {1954} (\bibinfo {year} {1971})},\ \bibinfo {note} {[Sov. Phys. JETP {\bf
  33}, 1053 (1971)]}\BibitemShut {NoStop}%
\bibitem [{\citenamefont {Suzuki}\ and\ \citenamefont
  {Murakami}(2025)}]{Suzuki2025}%
  \BibitemOpen
  \bibfield  {author} {\bibinfo {author} {\bibfnamefont {Y.}~\bibnamefont
  {Suzuki}}\ and\ \bibinfo {author} {\bibfnamefont {S.}~\bibnamefont
  {Murakami}},\ }\href@noop {} {\bibfield  {journal} {\bibinfo  {journal}
  {Phys. Rev. Lett.}\ }\textbf {\bibinfo {volume} {135}},\ \bibinfo {pages}
  {046301} (\bibinfo {year} {2025})}\BibitemShut {NoStop}%
\bibitem [{\citenamefont {Nakajima}(1973)}]{Nakajima1973}%
  \BibitemOpen
  \bibfield  {author} {\bibinfo {author} {\bibfnamefont {S.}~\bibnamefont
  {Nakajima}},\ }\href@noop {} {\bibfield  {journal} {\bibinfo  {journal}
  {Prog. Theor. Phys.}\ }\textbf {\bibinfo {volume} {50}},\ \bibinfo {pages}
  {1101} (\bibinfo {year} {1973})}\BibitemShut {NoStop}%
\bibitem [{\citenamefont {Sohn}\ \emph {et~al.}(2024)\citenamefont {Sohn},
  \citenamefont {Lee},\ and\ \citenamefont {Lee}}]{Sohn2024}%
  \BibitemOpen
  \bibfield  {author} {\bibinfo {author} {\bibfnamefont {J.}~\bibnamefont
  {Sohn}}, \bibinfo {author} {\bibfnamefont {J.~M.}\ \bibnamefont {Lee}},\ and\
  \bibinfo {author} {\bibfnamefont {H.-W.}\ \bibnamefont {Lee}},\ }\href@noop
  {} {\bibfield  {journal} {\bibinfo  {journal} {Phys. Rev. Lett.}\ }\textbf
  {\bibinfo {volume} {132}},\ \bibinfo {pages} {246301} (\bibinfo {year}
  {2024})}\BibitemShut {NoStop}%
\bibitem [{\citenamefont {Togawa}\ \emph {et~al.}(2019)\citenamefont {Togawa},
  \citenamefont {Kishine}, \citenamefont {Nosov}, \citenamefont {Koyama},
  \citenamefont {Paterson}, \citenamefont {McVitie}, \citenamefont {Kousaka},
  \citenamefont {Akimitsu}, \citenamefont {Ogata},\ and\ \citenamefont
  {Ovchinnikov}}]{Togawa2019}%
  \BibitemOpen
  \bibfield  {author} {\bibinfo {author} {\bibfnamefont {Y.}~\bibnamefont
  {Togawa}}, \bibinfo {author} {\bibfnamefont {J.}~\bibnamefont {Kishine}},
  \bibinfo {author} {\bibfnamefont {P.~A.}\ \bibnamefont {Nosov}}, \bibinfo
  {author} {\bibfnamefont {T.}~\bibnamefont {Koyama}}, \bibinfo {author}
  {\bibfnamefont {G.~W.}\ \bibnamefont {Paterson}}, \bibinfo {author}
  {\bibfnamefont {S.}~\bibnamefont {McVitie}}, \bibinfo {author} {\bibfnamefont
  {Y.}~\bibnamefont {Kousaka}}, \bibinfo {author} {\bibfnamefont
  {J.}~\bibnamefont {Akimitsu}}, \bibinfo {author} {\bibfnamefont
  {M.}~\bibnamefont {Ogata}},\ and\ \bibinfo {author} {\bibfnamefont {A.~S.}\
  \bibnamefont {Ovchinnikov}},\ }\href@noop {} {\bibfield  {journal} {\bibinfo
  {journal} {Phys. Rev. Lett.}\ }\textbf {\bibinfo {volume} {122}},\ \bibinfo
  {pages} {017204} (\bibinfo {year} {2019})}\BibitemShut {NoStop}%
\bibitem [{\citenamefont {Kanamori}(1963)}]{Kanamori1963}%
  \BibitemOpen
  \bibfield  {author} {\bibinfo {author} {\bibfnamefont {J.}~\bibnamefont
  {Kanamori}},\ }\href@noop {} {\bibfield  {journal} {\bibinfo  {journal}
  {Prog. Theor. Phys.}\ }\textbf {\bibinfo {volume} {30}},\ \bibinfo {pages}
  {275} (\bibinfo {year} {1963})}\BibitemShut {NoStop}%
\bibitem [{\citenamefont {Balkanski}\ \emph {et~al.}(1983)\citenamefont
  {Balkanski}, \citenamefont {Wallis},\ and\ \citenamefont
  {Haro}}]{Balkanski1983}%
  \BibitemOpen
  \bibfield  {author} {\bibinfo {author} {\bibfnamefont {M.}~\bibnamefont
  {Balkanski}}, \bibinfo {author} {\bibfnamefont {R.~F.}\ \bibnamefont
  {Wallis}},\ and\ \bibinfo {author} {\bibfnamefont {E.}~\bibnamefont {Haro}},\
  }\href@noop {} {\bibfield  {journal} {\bibinfo  {journal} {Phys. Rev. B}\
  }\textbf {\bibinfo {volume} {28}},\ \bibinfo {pages} {1928} (\bibinfo {year}
  {1983})}\BibitemShut {NoStop}%
\bibitem [{\citenamefont {Pauling}\ \emph {et~al.}(1951)\citenamefont
  {Pauling}, \citenamefont {Corey},\ and\ \citenamefont
  {Branson}}]{Pauling1951}%
  \BibitemOpen
  \bibfield  {author} {\bibinfo {author} {\bibfnamefont {L.}~\bibnamefont
  {Pauling}}, \bibinfo {author} {\bibfnamefont {R.~B.}\ \bibnamefont {Corey}},\
  and\ \bibinfo {author} {\bibfnamefont {H.~R.}\ \bibnamefont {Branson}},\
  }\href@noop {} {\bibfield  {journal} {\bibinfo  {journal} {Proc. Natl. Acad.
  Sci. U.S.A.}\ }\textbf {\bibinfo {volume} {37}},\ \bibinfo {pages} {205}
  (\bibinfo {year} {1951})}\BibitemShut {NoStop}%
\bibitem [{\citenamefont {Cochran}\ \emph {et~al.}(1952)\citenamefont
  {Cochran}, \citenamefont {Crick},\ and\ \citenamefont {Vand}}]{Cochran1952}%
  \BibitemOpen
  \bibfield  {author} {\bibinfo {author} {\bibfnamefont {W.}~\bibnamefont
  {Cochran}}, \bibinfo {author} {\bibfnamefont {F.~H.~C.}\ \bibnamefont
  {Crick}},\ and\ \bibinfo {author} {\bibfnamefont {V.}~\bibnamefont {Vand}},\
  }\href@noop {} {\bibfield  {journal} {\bibinfo  {journal} {Acta
  Crystallogr.}\ }\textbf {\bibinfo {volume} {5}},\ \bibinfo {pages} {581}
  (\bibinfo {year} {1952})}\BibitemShut {NoStop}%
\end{thebibliography}%

\section*{End Matter}
\textit{Appendix A: Shells and achiral screw axes}---Electrons must be able to carry orbital angular momentum as the phonons do, which requires a degenerate shell on each site; the minimal choice is the $p$ shell ($m_\ell=0,\pm1$), and since $L_\pm$ shifts $m_\ell$ by one step, only phonons with $m=\pm1$ (mod $n$) inject angular momentum at first order. \blue{Injection requires only that the pumping of $m=+1$ and $m=-1$ be unbalanced, as the helicity of $\sigma^+$ light fixes the direction in optical pumping.} For the $d$ shell $|m|\le2$ becomes bright, and the on-site interaction also needs $F^{(4)}$ (SM, Secs.~S1.1 and S4.1).
A screw axis with $p=n-p$ ($2_1$, $6_3$) is its own mirror image; the handedness then resides in the atoms at general positions (in CrNb$_3$S$_6$, the twisted S$_6$ octahedra around Cr), and the pump is written as
\begin{equation*}
J_1=\sum_\mu\ell_{\mu,\parallel}|\al_\mu|^2 ,
\end{equation*}
with $\ell_{\mu,\parallel}$ the axial component of $\ell_\mu$, which needs no CAM label and applies equally to $P2$ and $222$ (SM, Sec.~S9.1).

\nck{\textit{Appendix B: Temperature dependence and the nonlocal signal of NbSi$_2$}---A measured temperature dependence is described by \orange{one $\hbar\omega_\mu$ of the intrinsic decay and an electron side that is a sum of channels, constant (open window) or activated (window opening at $E_a$)} (SM, Sec.~S7); \rev{the Edelstein effect has no memory time at all,} and the signatures of SM Table~S1 distinguish the four cases.
For polycrystalline NbSi$_2$, the ratio $Y(T)/Y(300\,{\rm K})$ read from Fig.~4(b) of Ref.~\cite{Shishido2021} is flat at about $1.5$ below $\simeq45$~K and decreases above $\simeq60$~K. \orange{A constant electron side cannot follow both regimes. With a second, thermally activated electron channel,
\begin{equation*}
\gel=\gamma_0\Bigl[1+\frac{\gamma_1}{\gamma_0}e^{-E_a/k_BT}\Bigr],
\end{equation*}
Eq.~\eqref{eq:Pconv} reproduces the whole curve (rms deviation $0.023$) with $\hbar\omega_\mu=20$~meV fixed and
\begin{equation*}
\frac{\Gamma_0}{\gamma_0}\simeq0.6,\qquad\frac{\gamma_1}{\gamma_0}\simeq4,\qquad E_a\simeq32~\text{meV};
\end{equation*}
the knee at $35$--$60$~K is sharper than the model, and the data do not fix whether the plateau is of the accumulation or the response type.}
\rev{The whole temperature dependence is carried by $\Pconv$; the radiation and the readout contribute none} (SM, Sec.~S8 and Fig.~S5).}

\textit{Nonlocal detection}---The scheme describes the generation of the reservoir and its readout where the current flows; the comparison above assumes that the detector is reached by the spreading current. A signal far from the current path requires transport of the angular momentum. The anharmonic decay $\Gph$ removes the quanta from the resonant mode but conserves the CAM modulo $n$, so the imbalance between $m=+1$ and $m=-1$ is passed to the acoustic branches. 

In the acoustic branches the angular momentum relaxes by a mechanism of the Dyakonov--Perel type \cite{Dyakonov1971,Suzuki2025}: elastic scattering, which randomizes the wave vector and with it the quantization axis, slows the relaxation by motional narrowing, so that the diffusion length $\ell_{\rm AM}\sim v/\Omega$, with $\hbar\Omega(q)$ the longitudinal--transverse splitting of the acoustic branches, is independent of the disorder and grows for long-wavelength phonons, until anharmonic scattering takes over at high temperature \cite{Suzuki2025}. The two cases differ in the distance dependence: geometric and temperature independent for the spreading current, $e^{-x/\ell_{\rm AM}(T)}$ for the transported angular momentum, which is the nonlocal measurement proposed in Ref.~\cite{Suzuki2025}.

\textit{Appendix C: Stoner factors, paramagnon, and readout site}---The decomposition of Eq.~\eqref{eq:Hint} gives the Stoner constants
\begin{equation*}
I_L=F^{(0)}-\frac{F^{(2)}}{5},\qquad I_S=F^{(0)}+\frac{2F^{(2)}}{5},
\end{equation*}
which differ by the Hund coupling $3F^{(2)}/5$, so that $\mathcal{A}_S$ reaches its instability before $\mathcal{A}_L$ does.
Without SOC the channels do not mix and the RPA closes within each \cite{IKK1963},
\begin{equation*}
\mathcal{A}_{L,S}=\bigl[1-I_{L,S}\chi^0_{L,S}\bigr]^{-1};
\end{equation*}
the spin paramagnon of the readout is the odd-bubble chain that enhances the triplet channel \cite{Nakajima1973}, \rev{which near the instability reduces to one half of $\mathcal{A}_S$, the factor used throughout} (SM, Sec.~S5).

\nck{The orbital angular momentum is read out inside the medium or at an interface or electrode with atomic SOC (Au, Pt, Ni) within the orbital diffusion length $\ell_L$ \cite{Sohn2024}; the readout site changes only $\eta$, through $\lambda\mathcal{A}_S$, while the sign rule, $\tau_\mu$, and the orbital magnetization
\begin{equation*}
M_L=-\frac{\mu_B}{\hbar}\,\frac{\nu\tau_L}{V}\,\mathcal{A}_L\langle L_z\rangle_0
\end{equation*}
do not change. 

Near an instability, as in CrNb$_3$S$_6$ near $T_c$, $\mathcal{A}_S\gg1$ and the same $\langle L_z\rangle_0$ is read out as a much larger magnetization (SM, Sec.~S6.2). In thin lamellae of CrNb$_3$S$_6$ the helimagnetic stripe order survives up to about $140$~K, well above the bulk $T_c=123$~K \cite{Togawa2019}, which suggests that this compound stays close to its magnetic instability, and hence has a large $\mathcal{A}_S$, over a considerable range of temperatures.}

\clearpage
\setcounter{section}{0}
\setcounter{subsection}{0}
\setcounter{equation}{0}
\setcounter{figure}{0}
\setcounter{table}{0}
\renewcommand{\thesection}{S\arabic{section}}
\renewcommand{\thesubsection}{S\arabic{section}.\arabic{subsection}}
\renewcommand{\theequation}{S\arabic{equation}}
\renewcommand{\thefigure}{S\arabic{figure}}
\renewcommand{\thetable}{S\arabic{table}}
\allowdisplaybreaks
\setcounter{secnumdepth}{2}
\onecolumngrid
\begin{center}
{\large\bfseries Supplemental Material for\\
``Chirality-Induced Angular-Momentum Pumping by Moving Charges''}\\[4pt]
Jun-ichiro Kishine and Ron Naaman
\end{center}
\vspace{8pt}
\twocolumngrid

\noindent
This Supplemental Material gives the derivations and additional details for the scheme of the main text.
Figure~S1 summarizes the six steps, and Figure~S2 represents the theoretical scheme as Feynman diagrams of the pumping process. Part~I (Secs.~S1--S6) derives steps (1)--(6) of the main text in order; Part~II (Secs.~S7--S8) treats the temperature dependence and the analysis of the NbSi$_2$ data; Part~III (Sec.~S9) collects extensions; and Part~IV (Sec.~S10) gives the material map.
Notation: $\hbar$ is kept explicit, $c$ is the lattice constant along the axis, $m$ is the crystal angular momentum (CAM), $m_\ell$ the local orbital quantum number, and $\ell_\mu$ the mechanical angular momentum of mode $\mu$; $\boldsymbol u_\ell$ is the displacement of ion $\ell$ from its equilibrium position and $\boldsymbol\xi_{\mu,\ell}$ the displacement vector of mode $\mu$ on site $\ell$,
\begin{equation}
\boldsymbol u_\ell=\sum_\mu Q_\mu\boldsymbol\xi_{\mu,\ell}.
\end{equation}

\begin{figure*}[t!]
\centering
\includegraphics[width=0.9\textwidth]{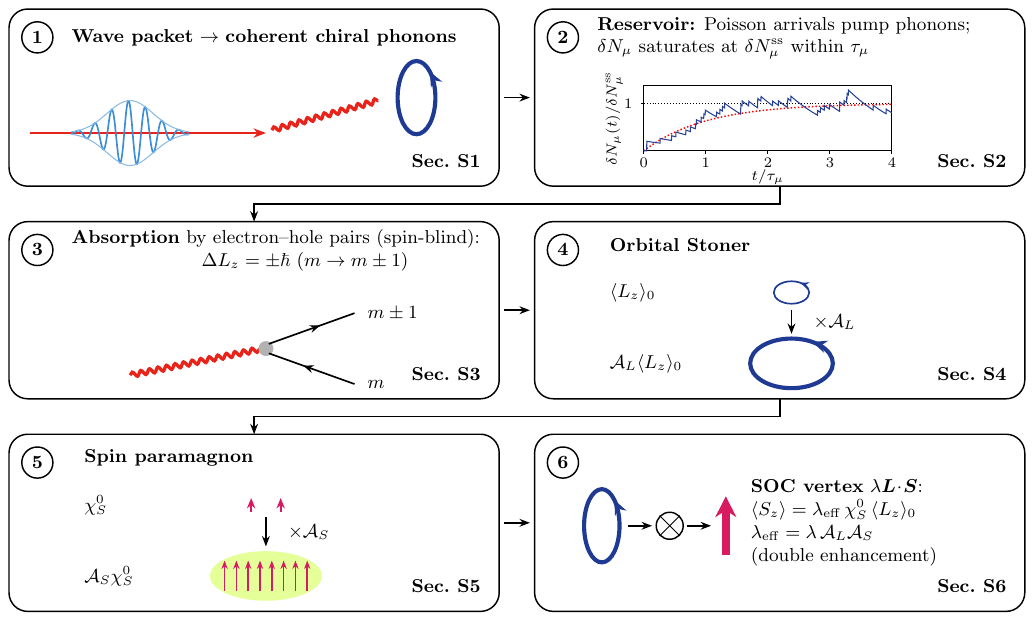}
\caption{The six steps of CAMP and the sections of this Supplemental Material that treat them.
(1) A wave packet radiates phase-matched chiral phonons coherently.
(2) Under randomly timed arrivals (ticks) the occupation $\delta N_\mu$ (blue, one realization) accumulates toward the steady state $\delta N^{\rm ss}_\mu$ (red, rate equation) with memory time $\tau_\mu$: coherent radiation by each electron, incoherent accumulation over electrons.
(3) Electron--hole pairs read out the reservoir through a spin-independent vertex that changes $m$ to $m\pm1$.
(4), (5) The orbital Stoner factor $\mathcal{A}_L$ and the spin paramagnon $\mathcal{A}_S$ enhance $\langle L_z\rangle_0$ and $\chi^0_S$, respectively, through the same on-site interaction.
(6) A single spin--orbit vertex joins them, $\langle S_z\rangle=\lambda_{\rm eff}\chi^0_S\langle L_z\rangle_0$ with $\lambda_{\rm eff}=\lambda \mathcal{A}_L\mathcal{A}_S$.}
\label{fig:S1}
\end{figure*}

\begin{figure*}[t!]
\centering
\includegraphics[width=0.9\textwidth]{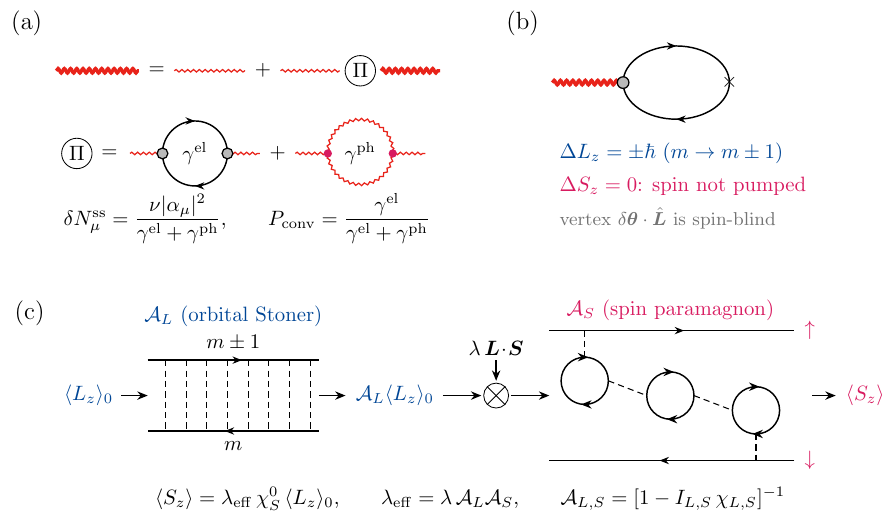}
\caption{Diagrams of the pumping process.
(a) Dyson equation of the driven phonon (thin and thick wavy lines: bare and dressed phonon). The self-energy $\Pi$ consists of the electron--hole bubble with electron--phonon vertices (gray dots), giving the electronic damping $\gel$, and the phonon loop with anharmonic vertices (red dots), giving the phononic damping $\Gph$; they fix the steady-state occupation $\delta N^{\rm ss}$ and the conversion probability $\Pconv$.
(b) Bare absorption of a stored phonon by an electron--hole pair. The gray dot is the rotation vertex $\delta\boldsymbol\theta\cdot\hat{\boldsymbol L}$, which transfers $\Delta L_z=\pm\hbar$ ($m\to m\pm1$) and does not depend on the spin; the cross is the $L_z$ insertion that reads out the transferred angular momentum.
(c) Double enhancement. Dashed lines are the on-site Coulomb interaction. The orbital Stoner ladder $\mathcal{A}_L$ dresses the $m\to m\pm1$ pairs; a single spin--orbit vertex $\lambda\boldsymbol L\cdot\boldsymbol S$ (crossed circle) feeds the result into the spin paramagnon $\mathcal{A}_S$, the chain of an odd number of bubbles of the same interaction between the $\uparrow$ and $\downarrow$ lines (the chain of Nakajima \cite{Nakajima1973} that enhances the triplet channel; its ease of enhancement is Hund's first rule, which enters the Casimir form of the on-site interaction with four times the weight of the second rule, Sec.~S5), giving $\langle S_z\rangle=\lambda_{\rm eff}\chi^0_S\langle L_z\rangle_0$ with $\lambda_{\rm eff}=\lambda \mathcal{A}_L\mathcal{A}_S$ and $\mathcal{A}_{L,S}=[1-I_{L,S}\chi_{L,S}]^{-1}$.}
\label{fig:S2}
\end{figure*}

\section*{Part I: Derivations of the six steps}

\section{Details on Step (1): wave-packet radiation}
This section derives step (1) of the main text; step (2) is treated in Sec.~S2.

\subsection{Structure}
The helix is a screw $\{C_n|\tau\}$ with $\tau=pc/n$; for the $3_1$ helix $\tau=c/3$ is the site spacing. Phonon modes are labeled by the screw momentum $\kappa=q+2\pi m/c$ and the crystal angular momentum (CAM) $m$; the rotational electron--phonon vertex is that of Refs.~\cite{Kishine2020,KatoKishine2023,Tateishi2026,Tateishi2025}. The rotation vertex shifts $m_\ell$ by one step, so only $m=\pm1$ (mod $n$) are bright at first order; the readable sectors are $|m|\le l_{\rm shell}$ ($|m|\le1$ for $p$ orbitals, $|m|\le2$ for $d$ orbitals, e.g.\ $m=\pm2$ in NbSi$_2$), and for large $n$ (the $\alpha$ helix, DNA) $m$ becomes effectively continuous. The interchange $p\leftrightarrow n-p$ maps a helix onto its enantiomer ($3_1\leftrightarrow3_2$, $4_1\leftrightarrow4_3$, $6_1\leftrightarrow6_5$, $6_2\leftrightarrow6_4$) and $m\to-m$, which reverses the sign of the pump; a screw axis with $p=n-p$ ($2_1$, $4_2$, $6_3$) is its own mirror image and carries no handedness by itself. A crystal with such an axis can nevertheless be chiral: in CrNb$_3$S$_6$ ($P6_322$, a Sohncke group that is its own enantiomorph) the handedness resides in the S atoms at general positions, which form around each Cr a twisted octahedron (site symmetry $32$ rather than $\bar3m$, S--Cr--S angle $179.6^\circ$) and thereby a chiral helical arrangement; the CAM of the $6_3$ axis does not label this handedness, so for such crystals the pump is written in the CAM-free form of Sec.~S9.1. The phonon dispersion of Fig.~1(b) is the closed-form solution of the $3\times3$ dynamical matrix of the $3_1$ helix \cite{Tateishi2025} with the illustrative force constants $\Omega_{xx}=\Omega_{yy}=78.4$, $\Omega_{zz}=98.0$, $\Omega_{xy}=49.0$, $\Omega_{yz}=\Omega_{xz}=29.4$ (arbitrary units), which are not material-specific.

\subsection{Wave-packet injection and forced oscillation}
Let $\hat R_z(\phi)$ denote the rotation by angle $\phi$ about the $z$ axis, acting on displacement vectors,
\begin{equation}
\hat R_z(\phi)=\begin{pmatrix}\cos\phi&-\sin\phi&0\\ \sin\phi&\cos\phi&0\\ 0&0&1\end{pmatrix} .
\end{equation}
The ion at site $\ell$ sits at
\begin{equation}
\begin{gathered}
\boldsymbol R_\ell=(\rho\cos\phi_\ell,\ \rho\sin\phi_\ell,\ z_\ell),\\
\phi_\ell=\frac{2\pi\ell}{3},\qquad z_\ell=\ell\tau ,
\end{gathered}
\end{equation}
where $\rho$ is the helix radius, the chain has $N$ sites, $\ell=0,\dots,N-1$, and length $L=N\tau$, so that the screw operation $\{C_3|\tau\}$ (a rotation by $2\pi/3$ about $z$ followed by a translation $\tau$ along $z$) maps site $\ell$ onto site $\ell+1$. The local (co-rotating) frame of site $\ell$ is obtained from that of site $0$ by the rotation $\hat R_z(\phi_\ell)$. The vector $\boldsymbol f_0(t)$ is the force that the charge exerts on the ion at site $0$, expressed in this frame, and $\boldsymbol\xi_{\mu,0}$ is the displacement pattern of mode $\mu$ on site $0$. The Bloch-type form of $\boldsymbol\xi_{\mu,\ell}$ below follows from the screw eigenvalue of mode $\mu$.

The normal modes are
\begin{equation}
\boldsymbol\xi_{\mu,\ell}=N^{-1/2}\,e^{-i\kappa_\mu z_\ell}\,\hat R_z(\phi_\ell)\,\boldsymbol\xi_{\mu,0},
\end{equation}
which are orthonormal,
\begin{equation}
\sum_\ell\boldsymbol\xi^*_{\mu,\ell}\cdot\boldsymbol\xi_{\mu',\ell}=\delta_{\mu\mu'} .
\end{equation}
The factor $e^{-i\kappa_\mu z_\ell}$ is the character of the screw $\{\hat R_z(2\pi/3)|\tau\}$,
\begin{equation}
\chi^{(\kappa)}(\hat S^\ell)=e^{-i\kappa z_\ell};
\end{equation}
the spatial phase of the phonon is thus $e^{-i\kappa z}$, and its positive-frequency component is
\begin{equation}
\boldsymbol u^{(+)}_\mu(z,t)\propto e^{-i(\kappa_\mu z+\omega_\mu t)}.
\end{equation}
Compared with a conventional propagating wave $e^{i(qz-\omega t)}$, its wave number is $q_{\rm prop}=-\kappa_\mu$.
The lattice displacement of ion $\ell$ is
\begin{equation}
\boldsymbol u_\ell(t)=\sum_\mu Q_\mu(t)\,\boldsymbol\xi_{\mu,\ell},
\end{equation}
with $Q_\mu$ the normal coordinate, and the ions obey
\begin{equation}
M\ddot{\boldsymbol u}_\ell=-\sum_{\ell'}K_{\ell\ell'}\boldsymbol u_{\ell'}+\boldsymbol f_\ell(t),
\end{equation}
where $K_{\ell\ell'}$ is the force-constant matrix and $\boldsymbol f_\ell$ the force exerted by the charge. For each mode,

\begin{equation}
\begin{gathered}
M\ddot Q_\mu+M\omega_\mu^2Q_\mu=F_\mu(t),\\
F_\mu(t)=\sum_\ell\boldsymbol\xi^*_{\mu,\ell}\cdot\boldsymbol f_\ell(t) ,
\end{gathered}
\end{equation}

A charge $Q$ moving along the axis with velocity $v>0$ reaches ion $\ell$ (charge $Ze$) at $t_\ell=z_\ell/v$ and exerts the screened Coulomb force
\begin{equation}
\boldsymbol f_\ell(t)=-\boldsymbol\nabla_\ell\Bigl[\frac{A\,e^{-k_s r_\ell(t)}}{r_\ell(t)}\Bigr],\qquad r_\ell(t)=|\boldsymbol R_\ell-vt\hat{\boldsymbol z}| ,
\end{equation}
with
\begin{equation}
A=\frac{QZe}{4\pi\varepsilon_0},\qquad k_s=\frac{1}{\lambda_s} .
\end{equation}
In the co-rotating frame,
\begin{equation}
\boldsymbol f_\ell(t)=\hat R_z(\phi_\ell)\,\boldsymbol f_0\!\Bigl(t-\frac{z_\ell}{v}\Bigr),
\end{equation}
so the projection on mode $\mu$ is
\begin{align}
F_\mu(t)&=\sum_\ell\boldsymbol\xi^*_{\mu,\ell}\cdot\boldsymbol f_\ell(t)\notag\\
&=\frac{1}{\sqrt N}\sum_{\ell=0}^{N-1}e^{i\kappa_\mu z_\ell}\,g_\mu\!\Bigl(t-\frac{z_\ell}{v}\Bigr),
\label{eq:Fsum}\\
g_\mu(t)&=\boldsymbol\xi^*_{\mu,0}\cdot\boldsymbol f_0(t).\notag
\end{align}
where $e^{i\kappa_\mu z_\ell}$ is the screw phase and $z_\ell/v$ the time delay; we used that $\hat R_z$ is real and orthogonal,
\begin{equation}
(\hat R_z\boldsymbol a)\cdot(\hat R_z\boldsymbol b)=\boldsymbol a\cdot\boldsymbol b.
\end{equation}
With
\begin{equation}
\tilde g_\mu(\omega)=\int dt\,e^{i\omega t}g_\mu(t),
\end{equation}
\begin{align}
\tilde F_\mu(\omega)&=\frac{\tilde g_\mu(\omega)}{\sqrt N}\,S_N(x_\mu),\\
S_N(x)&=\sum_{\ell=0}^{N-1}e^{i\ell x},\qquad x_\mu=\Bigl(\frac{\omega}{v}+\kappa_\mu\Bigr)\tau,
\end{align}
and
\begin{equation}
|S_N(x)|^2=\frac{\sin^2(Nx/2)}{\sin^2(x/2)}
\end{equation}
peaks at $x=2\pi j$ with height $N^2$ and width $\sim1/N$.

\begin{figure}[tb]
\centering
\includegraphics[width=0.95\columnwidth]{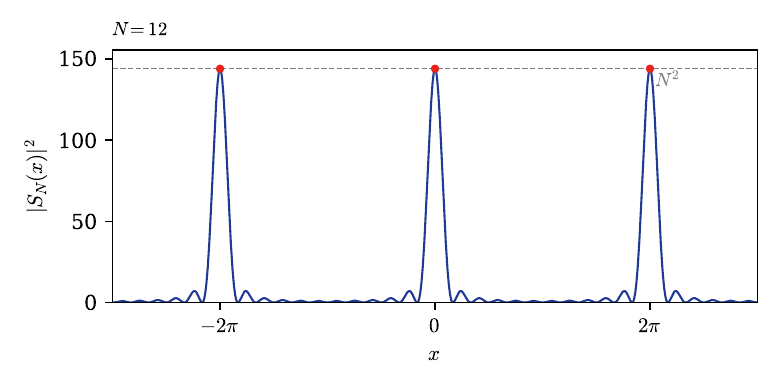}
\caption{The lattice sum $|S_N(x)|^2$ for $N=12$. The principal peaks at $x=2\pi j$ have height $N^2$ (dashed line), and the nearest zeros lie at distance $2\pi/N$ from each peak.
\label{fig:SN2}}
\end{figure}

The resonance condition is
\begin{equation}
\omega_\mu=v\,(G_j-\kappa_\mu),\qquad G_j=\frac{2\pi j}{\tau},\qquad j\in\mathbb Z ,
\label{eq:phase}
\end{equation}
where $j=0$ is the Cherenkov (normal) process and $j\neq0$ the Umklapp (Smith--Purcell \cite{SmithPurcell1953}) process. (For light, the delay between neighboring scatterers at distance $d$ is $\Delta t=d\cos\theta/c-d/v$, and constructive interference $\omega\Delta t=2\pi j$ with $j=0$ would require $v=c/\cos\theta\ge c$, which is impossible; only $j\neq0$ survives.) For the normal process, the electron makes the transition $\boldsymbol k_i\to\boldsymbol k_f$ with
\begin{equation}
\boldsymbol k_f-\boldsymbol k_i=q_{\rm prop}=-\kappa_\mu,\qquad\text{i.e.,}\quad \boldsymbol k_f=\boldsymbol k_i-\boldsymbol\kappa_\mu,
\end{equation}
and energy conservation
\begin{equation}
\varepsilon_{\boldsymbol k-\boldsymbol\kappa_\mu}-\varepsilon_{\boldsymbol k}\simeq-\hbar v_g^{\rm el}(\boldsymbol k)\kappa_\mu=\hbar\omega_\mu=-\hbar v\kappa_\mu
\end{equation}
requires
\begin{equation}
v_g^{\rm el}(\boldsymbol k)=v ,
\end{equation}
that is, the electron group velocity must match the injection velocity. No SOC enters this step.

A Bloch electron in a stationary state has
\begin{equation}
\langle\hat\rho_{\boldsymbol q}\rangle=0
\end{equation}
and emits phonons only incoherently, whereas the packet fixes the kick time of each ion and gives a coherent amplitude
\begin{equation}
\al_\mu\neq0 .
\end{equation}
Incoherent emission populates $m=\pm1$ equally for a time-reversal-symmetric distribution, so
\begin{equation}
\sum_\mu\ell_\mu\dN_\mu=0 ,
\end{equation}
and it is not included in the rate equation of Sec.~S2.2. A current-carrying distribution can make it asymmetric (the response-type Edelstein effect, Table~\ref{tab:fp}), which is distinct from the pump of this work. In three dimensions the transverse wave vector $\boldsymbol q_\perp$ turns each resonance into a surface, and the sum over modes becomes an integral over it weighted by the transverse spectrum of the pulse; roots with opposite $\ell_\mu$ partly cancel, and the single-resonance limit of Sec.~S3.1 is the case in which one root dominates.

The phonon Hamiltonian is
\begin{equation}
\begin{gathered}
\hat H_0=\sum_\mu\hbar\omega_\mu\hat a^\dagger_\mu\hat a_\mu,\qquad u^{\rm ZP}_\mu=\sqrt{\hbar/2M\omega_\mu},\\
\hat Q_\mu=u^{\rm ZP}_\mu(\hat a_\mu+\hat a^\dagger_\mu).
\end{gathered}
\end{equation}
The driven mode has
\begin{equation}
\hat H(t)=\hat H_0-\sum_\mu u^{\rm ZP}_\mu\bigl(\hat a^\dagger_\mu F_\mu(t)+\hat a_\mu F^*_\mu(t)\bigr) .
\end{equation}
The Heisenberg equation is linear,
\begin{equation}
i\hbar\,\dot{\hat a}_\mu=[\hat a_\mu,\hat H]=\hbar\omega_\mu\hat a_\mu-u^{\rm ZP}_\mu F_\mu(t).
\end{equation}
In the co-rotating form
\begin{equation}
\hat a_\mu(t)=e^{-i\omega_\mu t}\hat b_\mu(t)
\end{equation}
it reads
\begin{equation}
\dot{\hat b}_\mu=(iu^{\rm ZP}_\mu/\hbar)\,e^{i\omega_\mu t}F_\mu(t),
\end{equation}
so that

\begin{align}
\hat a_\mu(t)&=e^{-i\omega_\mu t}\bigl[\hat a_\mu(0)+\al_\mu(t)\bigr],\\
\al_\mu(t)&=\frac{iu^{\rm ZP}_\mu}{\hbar}\int_0^t dt'\,e^{i\omega_\mu t'}F_\mu(t')\notag\\
&=\frac{i}{\sqrt{2M\hbar\omega_\mu}}\int_0^t dt'\,e^{i\omega_\mu t'}F_\mu(t') .
\end{align}

One passage leaves the mode in a coherent state with
\begin{align}
\al_\mu&=\frac{iu^{\rm ZP}_\mu}{\hbar}\tilde F_\mu(\omega_\mu),\notag\\
|\al_\mu|^2&=\frac{|\tilde F_\mu(\omega_\mu)|^2}{2M\hbar\omega_\mu}=\frac{|\tilde g_\mu(\omega_\mu)|^2}{2M\hbar\omega_\mu N}\,|S_N(x_\mu)|^2 ,
\label{eq:alphaF}
\end{align}
so the whole velocity dependence is that of $\tilde g_\mu$. The potential
\begin{equation}
U=\frac{A\,e^{-k_s\sqrt{\rho^2+\zeta^2}}}{\sqrt{\rho^2+\zeta^2}},\qquad \zeta=z-vt ,
\end{equation}
with $\rho$ the impact parameter, satisfies
\begin{equation}
\begin{gathered}
\int d\zeta\,e^{-i\omega \zeta/v}\,U=2A\,K_0(\rho\beta),\\
\beta=\sqrt{\Bigl(\frac{\omega}{v}\Bigr)^2+k_s^2} ,
\end{gathered}
\end{equation}
so that its time transform is
\begin{equation}
\tilde U=\frac{2A}{v}K_0(\rho\beta).
\end{equation}
With
\begin{equation}
\tilde f_\rho=-\partial_\rho\tilde U,\qquad \tilde f_z=-i\frac{\omega}{v}\tilde U,\qquad K_0'=-K_1 ,
\end{equation}
we obtain
\begin{equation}
\tilde g_\mu(\omega)=\frac{2A}{v}\Bigl[\xi^*_{\mu,0,\rho}\,\beta K_1(\rho\beta)-i\,\xi^*_{\mu,0,z}\,\frac{\omega}{v}K_0(\rho\beta)\Bigr],
\label{eq:FT}
\end{equation}
where $\xi_{\mu,0,\rho}$ and $\xi_{\mu,0,z}$ are the radial and axial components of $\boldsymbol\xi_{\mu,0}$. For $k_s\rho\ll1$,
\begin{equation}
\beta\simeq\frac{\omega}{v},\qquad x\equiv\rho\beta=\frac{\omega\rho}{v},
\end{equation}
and the radial term dominates, so on resonance
\begin{equation}
|\al_\mu|^2=C_\mu\,x^4K_1^2(x),\qquad C_\mu=\frac{2A^2|\xi_{\mu,0,\rho}|^2}{M\hbar\rho^4\omega_\mu^3},
\label{eq:alphax}
\end{equation}
with $C_\mu$ containing the lattice-sum factor $|S_N|^2/N$ on resonance. Both $k_s\rho\gg1$ and $\rho\omega/v\gg1$ give
\begin{equation}
\rho\beta\gg1,\qquad |\tilde g_\mu|^2\propto\frac{\beta}{\rho v^2}\,e^{-2\rho\beta},
\end{equation}
so that the near field of the electron cannot excite that phonon effectively. In the regime $x\ll1$ (impulse approximation, $\Delta p=2A/\rho v$),
\begin{equation}
|\al_\mu|^2\to\frac{2A^2|\xi_{\mu,0,\rho}|^2}{M\hbar\rho^2\omega_\mu v^2}\propto v^{-2} ,
\end{equation}
and in the regime $x\gg1$,
\begin{equation}
\begin{gathered}
|\al_\mu|^2\to\frac{\pi}{2}C_\mu x^3e^{-2x}\propto u^{-3}e^{-2/u},\\
u\equiv\frac{v}{\rho\omega_\mu}=\frac1x ,
\end{gathered}
\end{equation}
The exponential is the adiabatic cutoff: the passage time $\Delta t\sim\rho/v$ exceeds the phonon period. For the dominant resonance, $\omega_\mu=\omega_*$, this is the variable $u=v/\rho\omega_*$ of Fig.~2(b) of the main text.

The phonon number grows by $\delta N_\mu=|\al_\mu|^2$, since
\begin{equation}
\langle\hat N_\mu(t)\rangle=\langle(\hat a^\dagger_\mu(0)+\al^*_\mu)(\hat a_\mu(0)+\al_\mu)\rangle\to|\al_\mu|^2
\end{equation}
for a mode initially empty. For $N\to\infty$, $|S_N(x)|^2\to2\pi N\sum_{j\in\mathbb Z}\delta(x-2\pi j)$, so
\begin{equation}
\begin{split}
|\al_\mu|^2\simeq{}&\frac{|\tilde g_\mu(\omega_\mu)|^2}{2M\hbar\omega_\mu N}\\
&\times2\pi N\sum_{j\in\mathbb Z}\delta\bigl[(\omega_\mu/v+\kappa_\mu)\tau-2\pi j\bigr] ,
\end{split}
\end{equation}
and near a resonance
\begin{equation}
x(\kappa)=\tau[\omega(\kappa)/v+\kappa]\simeq\tau(1-v_g^{\rm ph}/v)(\kappa-\kappa_0),
\end{equation}
$dx/d\kappa=\tau(1-v_g^{\rm ph}/v)$, which is the Jacobian of the resonance; here
\begin{equation}
v_g^{\rm ph}=\partial\omega/\partial q_{\rm prop}=-\partial\omega/\partial\kappa
\end{equation}
is the phonon group velocity, and the density of resonant roots carries $|dx/d\kappa|$.

The phonon angular momentum is
\begin{equation}
J^{\rm ph}_z=\sum_\ell M(\boldsymbol u_\ell\times\dot{\boldsymbol u}_\ell)_z
\end{equation}
with
\begin{equation}
\boldsymbol u_\ell=\sum_\mu u^{\rm ZP}_\mu(\hat a_\mu\boldsymbol\xi_{\mu,\ell}+\hat a^\dagger_\mu\boldsymbol\xi^*_{\mu,\ell}),
\end{equation}
which gives
\begin{equation}
\langle J^{\rm ph}_z\rangle=-i\hbar\sum_\mu(N_\mu+\tfrac12)\sum_\ell(\boldsymbol\xi^*_{\mu,\ell}\times\boldsymbol\xi_{\mu,\ell})_z.
\end{equation}
With $N_\mu=N^{(0)}_\mu+\delta N_\mu$ the change is
\begin{equation}
\delta\langle J^{\rm ph}_z\rangle=\sum_\mu\ell_\mu|\al_\mu|^2.
\end{equation}

The angular momentum per electron is
\begin{equation}
J_1=\sum_\mu\ell_\mu|\al_\mu|^2 ,
\end{equation}
where $\ell_\mu$ is the mechanical angular momentum of mode $\mu$,
\begin{equation}
\ell_\mu=-i\hbar\sum_\ell\bigl(\boldsymbol\xi^*_{\mu,\ell}\times\boldsymbol\xi_{\mu,\ell}\bigr)_z .
\end{equation}
For a fixed mode, $\ell_\mu$ is set by the eigenvector and does not depend on $v$. The net value of $J_1$ is the asymmetry between the sectors $m=+1$ and $m=-1$, which have opposite $\ell_\mu$ and equal $|\xi_{\mu,0,\rho}|^2$; it arises because the two sectors resonate at frequencies, and with Jacobians $\tau(1-v_g^{\rm ph}/v)$, that differ by the chiral splitting of the branches. To lowest order in the splitting the asymmetry is proportional to the retardation between neighboring ions, $\omega_\mu\tau/v\propto x$, whereas the function multiplying it depends on the branches. We take the asymmetry as $\propto x$ with the common envelope $x^4K_1^2(x)$; this is the model used in Fig.~2(b) of the main text. Hence
\begin{equation}
\begin{gathered}
J_1\propto x^5K_1^2(x)\;\longrightarrow\\
\begin{cases}x^3\propto v^{-3}&(x\ll1),\\[2pt] \tfrac\pi2x^4e^{-2x}=\tfrac\pi2u^{-4}e^{-2/u}&(x\gg1).\end{cases}
\end{gathered}
\label{eq:J1x}
\end{equation}
Maximizing $x^kK_1^2$ gives $K_1/K_0=2x/(k-2)$: $x\simeq1.33$ for $|\al_\mu|^2$ ($k=4$) and $x\simeq1.87$ for $J_1$ ($k=5$), i.e.\ $u^*=v^*/\rho\omega_*\simeq0.54$, the optimum of Fig.~2(b). For $\rho\simeq1$--$2$~\AA\ and $\hbar\omega_\mu\sim10$~meV this gives $v^*\sim10^3$~m/s, of the order of the sound velocity. $J_1$ changes sign with the enantiomer and with $v\to-v$, and vanishes in any achiral structure, where the degenerate modes $(\kappa,\ell)$ and $(\kappa,-\ell)$ cancel. The screw symmetry lifts this degeneracy (Sec.~S1.1). Chirality thus enters at this stage: the direction of injection, a polar vector, selects the sign of the radiated angular momentum, an axial vector.

\section{Details on Step (2): accumulation and rate equation}

\subsection{Poisson averaging}

The electrons arrive at the instants $t_p$ as a Poisson process with mean rate $\nu=I/e$, so that the arrival density
\begin{equation}
n(t)=\sum_p\delta(t-t_p),\qquad\langle n(t)\rangle=\nu,\qquad I=e\nu ,
\end{equation}
has the shot-noise spectrum
\begin{align}
S_{\rm arr}(\omega)&\equiv\int_{-\infty}^{\infty}\mathrm{d}\tau\,e^{i\omega\tau}\langle\delta n(t+\tau)\,\delta n(t)\rangle\notag\\
&=\nu ,\\
\langle\delta n(t)\,\delta n(t')\rangle&=\nu\,\delta(t-t') .\notag
\end{align}

The electron arriving at $t_p$ leaves the amplitude $\al_\mu e^{i\omega_\mu t_p}$ in mode $\mu$, where the phase $e^{i\omega_\mu t_p}$ is the delay, so that after the time $T$
\begin{align}
A_\mu(T)&=\al_\mu\sum_{t_p\in[0,T]}e^{i\omega_\mu t_p},\\
|A_\mu(T)|^2&=|\al_\mu|^2\sum_{p,p'}e^{i\omega_\mu(t_p-t_{p'})} .
\end{align}

For a (possibly complex) function $f(t)$,
\begin{equation}
\sum_{t_p\in[0,T]}f(t_p)=\int_0^T n(t)f(t)\,\mathrm{d}t.
\end{equation}
With
\begin{equation}
\langle n(t)\,n(t')\rangle=\nu^2+\nu\,\delta(t-t'),
\end{equation}
\begin{align}
\Bigl\langle\sum_{t_p\in[0,T]}f(t_p)\Bigr\rangle&=\nu\int_0^T f(t)\,\mathrm{d}t, \\
\Bigl\langle\Bigl|\sum_{t_p\in[0,T]}f(t_p)\Bigr|^2\Bigr\rangle&=\nu\int_0^T|f(t)|^2\,\mathrm{d}t\notag\\
&\quad+\Bigl|\nu\int_0^T f(t)\,\mathrm{d}t\Bigr|^2 .
\label{eq:campbell}
\end{align}

Take $f(t)=e^{i\omega_\mu t}$ with $\omega=\omega_\mu$. Then
\begin{equation}
\int_0^T e^{i\omega t}\,\mathrm{d}t=(e^{i\omega T}-1)/(i\omega),
\end{equation}
and
\begin{align}
\Bigl\langle\Bigl|\sum_{t_p\in[0,T]}e^{i\omega t_p}\Bigr|^2\Bigr\rangle&=\nu T+\nu^2\Bigl|\frac{e^{i\omega T}-1}{i\omega}\Bigr|^2\notag\\
&\le\nu T+\frac{4\nu^2}{\omega^2}.
\end{align}
For $\omega\neq0$ the second term does not grow with $T$, so
\begin{equation}
\Bigl\langle\Bigl|\sum_{t_p}e^{i\omega t_p}\Bigr|^2\Bigr\rangle\;\to\;\nu T\qquad(T\to\infty).
\end{equation}
The second term in Eq.~\eqref{eq:campbell} is the square of the mean, which is bounded here: the phase factors $e^{i\omega t_p}$ of different electrons do not add coherently in power. For $\omega=0$ the mean term would grow as $\nu^2T^2$ and dominate, so the argument requires $\omega\neq0$, which holds for the resonant modes.
Since
\begin{equation}
A_\mu(T)=\al_\mu\sum_{t_p}e^{i\omega_\mu t_p}
\end{equation}
(see above), this gives
\begin{equation}
\lim_{T\to\infty}\frac{\langle|A_\mu(T)|^2\rangle}{T}=\nu|\al_\mu|^2\equiv R^{\rm pump}_\mu .
\end{equation}
Thus the radiation by each electron is coherent, whereas the accumulation over electrons is incoherent. Coherence is needed only within one passage, where it sets $\al_\mu$ and the asymmetry between $m=+1$ and $m=-1$; the later steps use only the occupations.

The excess occupation $\dN_\mu$ over thermal equilibrium, and the angular momentum $\ell_\mu\dN_\mu$ it carries, form the reservoir. Let $\dN^{(m)}\equiv\sum_{\mu\in m}\dN_\mu$ denote the occupation of the CAM sector $m$. Time-reversal-odd observables appear only through $\dN^{(+1)}-\dN^{(-1)}$, and time-reversal-even observables only in proportion to $\dN_\mu$.

\subsection{Rate equation}
The phonon line is dressed as in Fig.~S2(a): $D=D_0+D_0\Pi D$ with $\Pi=\Pi^{\rm el}+\Pi^{\rm ph}$, and
\begin{equation}
\gamma=-2\,{\rm Im}\,\Pi(\omega_\mu)/\hbar.
\end{equation}
The occupation obeys
\begin{equation}
\frac{d\dN_\mu}{dt}=R^{\rm pump}_\mu-(\gel_\mu+\Gph_\mu)\dN_\mu ,
\end{equation}
with the steady state
\begin{equation}
\dN^{\rm ss}_\mu=R^{\rm pump}_\mu/(\gel_\mu+\Gph_\mu),
\end{equation}
the memory time $\tau_\mu=1/(\gel_\mu+\Gph_\mu)$, and the conversion probability
\begin{equation}
\Pconv=\gel_\mu/(\gel_\mu+\Gph_\mu).
\end{equation}

Figure~2(a) of the main text compares one stochastic realization (arrivals drawn from a Poisson process; a $3_1$ helix of 126 sites, $\nu\tau_\mu=20$) with the solution of the rate equation.

\section{Details on Step (3): absorption and the orbital transfer}

\subsection{Golden rule and rotation vertex}
The electronic damping of mode $\mu$ is given by the golden rule,
\begin{equation}
\gel_\mu=\frac{2\pi}{\hbar}\sum_{i,f}|M_{fi}|^2\bigl[f(\varepsilon_i)-f(\varepsilon_f)\bigr]\delta(\varepsilon_f-\varepsilon_i-\hbar\omega_\mu),
\end{equation}
where $M_{fi}=\langle f|\delta H_\mu|i\rangle$ is the one-phonon matrix element of the electron--phonon coupling and the occupation factor defines the Pauli window: in a metal it is $\simeq\hbar\omega_\mu(-\partial f/\partial\varepsilon)$ near the Fermi surface, in a molecule it is the transition between occupied and empty orbitals across the gap.

The coupling is the rotational electron--phonon vertex of Ref.~\cite{Tateishi2026}.
The $p$ orbitals of site $l$ co-rotate with the local frame, so that the fermion operators of the laboratory and of the local frame are related by $\hat U_l=\exp[-\tfrac{i}{\hbar}\boldsymbol\theta_l\cdot\hat{\boldsymbol L}]$, where $\boldsymbol\theta_l$ contains the global helical rotation and the phonon-induced local rotation
\begin{equation}
\delta\boldsymbol\theta_l\simeq\frac{3}{2c}\,\boldsymbol e^3\times(\boldsymbol u_{l+1}-\boldsymbol u_l),
\end{equation}
defined on the discrete lattice from the relative displacement of neighboring sites ($c=3\tau$ is the lattice constant), which is valid up to the Brillouin-zone boundary.
Expanding $\hat U_l$ to first order in $\delta\boldsymbol\theta_l$ and inserting it into the tight-binding Hamiltonian $-\sum_{\langle l,l'\rangle}\hat c^\dagger_l\,t(\boldsymbol r_l-\boldsymbol r_{l'})\,\hat c_{l'}$ gives, besides the usual displacement coupling, the term
\begin{equation}
\hat H^{\rm (rot)}=-\frac{i}{\hbar}\sum_{\langle l,l'\rangle}\Bigl\{\hat c^\dagger_l\,(\delta\boldsymbol\theta_l\cdot\hat{\boldsymbol L})\,t^0\,\hat c_{l'}-\hat c^\dagger_l\,t^0\,(\delta\boldsymbol\theta_{l'}\cdot\hat{\boldsymbol L})\,\hat c_{l'}\Bigr\},
\end{equation}
with $t^0$ the equilibrium hopping matrix between the $p$ orbitals of the two sites.
Since $\delta\boldsymbol\theta_l\perp\boldsymbol e^3$, $\delta\boldsymbol\theta_l\cdot\hat{\boldsymbol L}=\hat L^{(-)}\delta\theta^{(+)}_l+\hat L^{(+)}\delta\theta^{(-)}_l$, so that absorbing the circular component $m_s=\pm1$ of the displacement changes the electronic $m_\ell$ to $m_\ell-m_s$, with crystal momentum and CAM conserved.
The vertex vanishes for $s$ electrons, does not depend on the spin, and its CAM selection rule is that of the screw; this is the second stage at which chirality enters, the first being the radiation (Sec.~S1.2), and the enhancement factors (Secs.~S4 and S5) and the SOC readout (Sec.~S6) only propagate the sign fixed at these two stages.
The matrix element $M_{fi}$ of the golden rule is that of $\hat H^{\rm (rot)}$, of order $t^0\,(\hbar/M\omega_\mu)^{1/2}/c$, with $(\hbar/M\omega_\mu)^{1/2}/c\sim10^{-2}$ \cite{Tateishi2026}.
The orbital transfer per absorption, $\langle\Delta L_z\rangle_\mu$, is the ratio of the $L_z$-weighted to the bare absorption rate; it is not the mechanical $\ell_\mu$ of the phonon but the difference between the orbital textures $\langle L_z\rangle$ of the final and initial bands at the string endpoints.
Injection requires only that the pumping of $m=+1$ and $m=-1$ be unbalanced, with the dominant sign fixed by the chirality, as the helicity of $\sigma^+$ light fixes the direction in optical pumping.

\begin{figure}[tb]
\centering
\includegraphics[width=0.95\columnwidth]{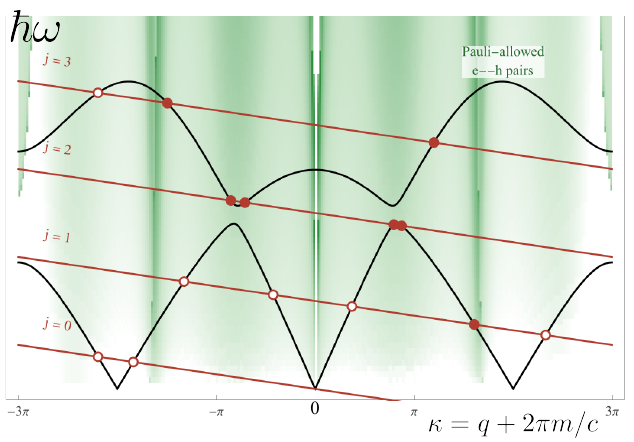}
\caption{Pauli window on the phonon dispersion (one-dimensional model, $\kappa=q+2\pi m/c$, $m=0,\pm1$).
Black: the two lowest of the three phonon branches of the $3_1$ helix of Sec.~S1.1 (the highest branch is omitted).
Red lines: the phase-matching conditions $\hbar\omega=v(G_j-\kappa)$, $j=0,\dots,3$, for a source of velocity $v$; their intersections with the branches (circles) are the resonances.
Green: the weight of Pauli-allowed electron--hole pairs that can absorb a phonon $(\kappa,\omega)$, computed for the lone-pair band of the $p$-orbital tight-binding model of Ref.~\cite{Tateishi2026} with scaled-down tellurium parameters, $E_F$ in the band, $T=300$~K (logarithmic shading; white: no weight).
Filled circles are resonances inside the window, which are absorbed; open circles lie outside it and are not.
Which resonances count is decided by the band structure and the position of $E_F$; in three dimensions the transverse momentum widens the window (Sec.~S3.1).}
\label{fig:pauli}
\end{figure}

Figure~\ref{fig:pauli} draws the window on the phonon dispersion of Fig.~1(b) for the lone-pair band of the model of Ref.~\cite{Tateishi2026} with $E_F$ in the band.
A phonon of screw momentum $\kappa$ and frequency $\omega$ can be absorbed only if the band structure has an occupied state $\kappa_e$ and an empty state $\kappa_e+\kappa$ separated by $\hbar\omega$; the resonances that lie in the window are absorbed (filled circles) and those outside it are not (open circles), so both the position of $E_F$ and the band structure decide which resonances count.
When $E_F$ lies in a gap $\Delta\gg\hbar\omega$, every resonance is blocked and $\gel$ is activated (Sec.~S7).
The electron energy scale of Fig.~\ref{fig:pauli} is halved relative to the tellurium parameters of Ref.~\cite{Tateishi2026}, and the figure is drawn for illustration; the phonon bands are those of Sec.~S1.1.

The window of Fig.~\ref{fig:pauli}, like that of the real tellurium band, is that of a one-dimensional band and is the most restrictive case.
For $\hbar\omega$ much smaller than the bandwidth, the condition
\begin{equation}
\varepsilon(\kappa_e+\kappa)-\varepsilon(\kappa_e)=\hbar\omega
\end{equation}
has isolated roots, each of which reads the window height $W=f(\varepsilon)-f(\varepsilon+\hbar\omega)$; at the Fermi level
\begin{equation}
W=\tfrac12\tanh(\hbar\omega/2k_BT),
\end{equation}
which is $0.11$ for $\hbar\omega=11.7$~meV at $300$~K and falls as $1/T$ at high temperature, so the window is a set of narrow stripes in $(\kappa,\omega)$.
In three dimensions the transverse momentum $\boldsymbol k_\perp$ is an additional free variable: for every $(\kappa,\omega)$ the condition is satisfied along a curve of the Fermi surface, and the sum over states gives $\hbar\omega$ times a density-of-states factor, independent of temperature for a smooth density of states.
The window is then open over broad ranges of $\kappa$ and most resonances lie inside it.
Accordingly, the electron side of the three-dimensional metals NbSi$_2$, TaSi$_2$, and CrNb$_3$S$_6$ contains a constant channel (Secs.~S7 and S10), whereas the one-dimensional Te band is the restrictive case.

When a single resonance dominates the absorption, the electron--phonon vertex, the Pauli factor, and the resonance Jacobian are common to the $L_z$-weighted and the bare rates and cancel in $\langle\Delta L_z\rangle_\mu$; for $\Pconv\to1$ the yield $Y$ of the main text is then fixed by the number of phonons radiated per electron and the orbital texture of the bands alone, independent of the coupling strength and of SOC, and the coupling survives only in $\Pconv$.

\section{Details on Step (4): orbital Stoner factor}

\subsection{On-site interaction of the $p$ shell}
The three $p$ orbitals are eigenstates $p_m$ of $L_z$, $m=+1,0,-1$. Their Coulomb and exchange integrals,
\begin{equation}
U_{mm'}=\langle mm'|V|mm'\rangle,\qquad J_{mm'}=\langle mm'|V|m'm\rangle,
\end{equation}
follow from the Slater integrals $F^{(0)}$ and $F^{(2)}$ through the Gaunt coefficients:
\begin{align}
U_{10}&=F^{(0)}-\tfrac{2}{25}F^{(2)}, & J_{10}&=\tfrac{3}{25}F^{(2)},\\
U_{11}&=U_{1,-1}=F^{(0)}+\tfrac{1}{25}F^{(2)}, & J_{1,-1}&=\tfrac{6}{25}F^{(2)},\\
U_{00}&=F^{(0)}+\tfrac{4}{25}F^{(2)}.
\end{align}

Two electrons in the $p$ shell form only three terms, ${}^3P$, ${}^1D$, and ${}^1S$. The state $|1\!\uparrow,0\!\uparrow\rangle$ has $L_z=1$ and $S_z=1$ and belongs only to ${}^3P$, so
\begin{equation}
E({}^3P)=U_{10}-J_{10}=F^{(0)}-\tfrac{F^{(2)}}5 .
\end{equation}
The state $|1\!\uparrow,1\!\downarrow\rangle$ has $L_z=2$ and belongs only to ${}^1D$, so
\begin{equation}
E({}^1D)=U_{11}=F^{(0)}+\tfrac{F^{(2)}}{25} .
\end{equation}
The three $L_z=S_z=0$ determinants span all three terms, and the trace of the $3\times3$ block gives
\begin{equation}
E({}^3P)+E({}^1D)+E({}^1S)=2U_{1,-1}+U_{00},
\end{equation}
hence
\begin{equation}
E({}^1S)=F^{(0)}+\tfrac{2F^{(2)}}5 .
\end{equation}

Since each pair of electrons carries a $\boldsymbol l_i\cdot\boldsymbol l_j$ and a $\boldsymbol s_i\cdot\boldsymbol s_j$ fixed by its term, the pair interaction can be written as
\begin{equation}
h_{ij}=\alpha+\beta\,\boldsymbol l_i\cdot\boldsymbol l_j+\gamma\,\boldsymbol s_i\cdot\boldsymbol s_j ,
\end{equation}
where, for ${}^3P$, ${}^1D$, and ${}^1S$ respectively,
\begin{equation}
\boldsymbol l_i\cdot\boldsymbol l_j=-1,\;+1,\;-2,\qquad \boldsymbol s_i\cdot\boldsymbol s_j=+\tfrac14,\;-\tfrac34,\;-\tfrac34 .
\end{equation}
Matching the three energies gives
\begin{align}
\beta&=\frac{E({}^1D)-E({}^1S)}3=-\frac{3F^{(2)}}{25},\\
\gamma&=E({}^3P)-E({}^1D)+2\beta=-\frac{12F^{(2)}}{25},\\
\alpha&=E({}^1D)-\beta+\tfrac34\gamma=F^{(0)}-\frac{F^{(2)}}5 .
\label{eq:abc}
\end{align}
This is exact for the $p$ shell because three energies are matched by three constants.

For $H_{\rm int}=\sum_{i<j}h_{ij}$ we use the total
\begin{equation}
\boldsymbol L=\sum_i\boldsymbol l_i,\qquad \boldsymbol S=\sum_i\boldsymbol s_i,
\end{equation}
with $\boldsymbol l_i^2=2$ and $\boldsymbol s_i^2=\tfrac34$:
\begin{align}
\sum_{i<j}1&=\frac{N(N-1)}2,\\
\sum_{i<j}\boldsymbol l_i\cdot\boldsymbol l_j&=\frac12(\boldsymbol L^2-2N),\\
\sum_{i<j}\boldsymbol s_i\cdot\boldsymbol s_j&=\frac12\Bigl(\boldsymbol S^2-\frac{3N}4\Bigr).
\end{align}
Hence
\begin{equation}
H_{\rm int}=\tfrac\alpha2N(N-1)+\tfrac\beta2(\boldsymbol L^2-2N)+\tfrac\gamma2\Bigl(\boldsymbol S^2-\tfrac{3N}4\Bigr),
\end{equation}
and inserting Eq.~\eqref{eq:abc} gives
\begin{align}
H_{\rm int}&=\tfrac12\Bigl(F^{(0)}-\tfrac{F^{(2)}}5\Bigr)\hat N(\hat N-1)+\tfrac3{10}F^{(2)}\hat N\notag\\
&\quad-\tfrac3{50}F^{(2)}\hat{\boldsymbol L}^2-\tfrac6{25}F^{(2)}\hat{\boldsymbol S}^2 ,
\label{eq:HintSM}
\end{align}
where the linear term follows from
\begin{equation}
-\beta-\tfrac38\gamma=\tfrac3{10}F^{(2)} .
\end{equation}
The single integral $F^{(2)}$ lowers the energy for large $\hat{\boldsymbol S}^2$ and for large $\hat{\boldsymbol L}^2$ alike, with coefficients in the ratio $4{:}1$; this is the operator form of Hund's first and second rules.

For the $d$ shell the on-site interaction also contains $F^{(4)}$, and the decomposition above is no longer exact; after spherical averaging, the Kanamori form \cite{Kanamori1963}
\begin{equation}
\frac{U-3J}2\hat N(\hat N-1)-2J\hat{\boldsymbol S}^2-\frac J2\hat{\boldsymbol L}^2
\end{equation}
is recovered.

\subsection{Stoner constants}
In the Hartree--Fock energy the Coulomb integral acts on all pairs, and the exchange integral lowers the energy of parallel-spin pairs. Polarizing the spin equally in the three orbitals gives the spin Stoner constant as the shell-averaged exchange integral,
\begin{equation}
I_S=F^{(0)}+\frac{2F^{(2)}}5 ,
\end{equation}
whereas the orbital polarization $m_L$ of the doublet $m=\pm1$ gives the orbital Stoner constant
\begin{equation}
I_L=F^{(0)}-\frac{F^{(2)}}5 .
\end{equation}
Both enter the Hubbard form of the polarization energy,
\begin{equation}
E_{\rm pol}=-\frac I4\,m^2 ,
\end{equation}
and the enhancement factors are
\begin{equation}
\mathcal{A}_{L,S}=\bigl[1-I_{L,S}\chi^0_{L,S}\bigr]^{-1}\qquad(\text{Sec.~S5}),
\end{equation}
where $\chi^0_S$ is the bare spin susceptibility per orbital and $\chi^0_L$ the bare orbital susceptibility of the doublet per spin direction. The difference
\begin{equation}
I_S-I_L=\frac{3F^{(2)}}5
\end{equation}
is the Hund coupling; it favors the spin channel, which therefore reaches its instability first.

In a molecule the Stoner denominator becomes the Hund-multiplet splitting of the excited configuration: the triplet exciton lies lower by $2K\propto F^{(2)}$, and the state with higher $L$ lies lower by the second rule.
The enhancement is then set by the fluctuations of the spin and orbital polarization of the injected configuration, and we expect it to become large when the configurations coupled by the SOC vertex approach degeneracy, so that a large $\mathcal{A}_S$ need not be restricted to metals. In a molecular junction the metallic electrode adds its own enhancement at the detection stage.

\section{Details on Step (5): shared origin and spin paramagnon}

Step (5) rests on the form of the on-site interaction, Eq.~\eqref{eq:HintSM}.
The coefficient of the density term,
\begin{equation}I_L=F^{(0)}-\tfrac{F^{(2)}}5\end{equation}
is the effective orbital repulsion of the shell (Sec.~S4.2).
The single integral $F^{(2)}$, the quadrupole (orientation-dependent) part of the Coulomb repulsion, multiplies $-\boldsymbol L^2$ and $-\boldsymbol S^2$ with the same sign: the energy is lowered by a large $L$ (Hund's second rule) and by a large $S$ (first rule), so the orbital and the spin ferromagnetic enhancements are consequences of one integral (spin--orbit coupling is a separate interaction).
The two coefficients are not equal,
\begin{equation}\tfrac{6}{25}F^{(2)}\ \text{for}\ \hat{\boldsymbol S}^2\qquad\text{against}\qquad\tfrac{3}{50}F^{(2)}\ \text{for}\ \hat{\boldsymbol L}^2,\end{equation}
a factor of four.
This is Hund's rules in operator form: the first rule (maximal $S$) precedes the second (maximal $L$) because the same integral lowers the energy four times more per unit of $\hat{\boldsymbol S}^2$ than per unit of $\hat{\boldsymbol L}^2$.
[With the spin written as $\hat{\boldsymbol\sigma}=2\hat{\boldsymbol S}$ the two coefficients coincide,
\begin{equation}-\tfrac{3}{50}F^{(2)}\bigl(\hat{\boldsymbol L}^2+\hat{\boldsymbol\sigma}^2\bigr);\end{equation}
the factor four is the $\hbar/2$ of the spin.]
The same hierarchy appears in the Stoner constants of Sec.~S4.2,
\begin{equation}I_S-I_L=\tfrac{3F^{(2)}}5,\end{equation}
for the same bare susceptibility the spin channel is closer to its instability than the orbital channel, and $\mathcal{A}_S$ grows before $\mathcal{A}_L$.
Accordingly, the triplet channel is the one that is easily enhanced, and the odd-bubble chain below, which enhances the triplet channel, is the relevant paramagnon of the readout.

Without SOC the on-site interaction is invariant under spin rotations and under orbital rotations separately, so the two channels do not mix and the random-phase approximation (RPA) closes within each channel.
We derive the RPA susceptibility for the spin channel; the orbital channel is isomorphic.

Let $m=n_\uparrow-n_\downarrow$ be the spin polarization per site and orbital, and let an external field $h$ couple to it through
\begin{equation}
H_h=-h\,\hat S_z=-\frac{h}{2}\,m .
\end{equation}
Without interaction the polarization responds linearly,
\begin{equation}
m=\chi^0_S\,h ,
\end{equation}
which defines the bare susceptibility $\chi^0_S$; for a band with density of states $\rho$ per spin and orbital at $E_F$, $\chi^0_S=\rho$.
The interaction enters through the polarization energy of Sec.~S4.2,
\begin{equation}
E_{\rm pol}=-\frac{I_S}{4}\,m^2 .
\end{equation}
In the mean-field (Hartree--Fock) treatment of this term, $m^2\to2m\langle m\rangle-\langle m\rangle^2$, the interaction acts on the polarization as a molecular field,
\begin{equation}
E_{\rm pol}\to-\frac{I_S}{2}\,\langle m\rangle\,m+\text{const},
\end{equation}
which has the same form as $H_h$ with $h\to I_S\langle m\rangle$.
The electrons therefore respond with the bare susceptibility to the total field,
\begin{equation}
m=\chi^0_S\bigl(h+I_S\,m\bigr),
\end{equation}
and solving this self-consistency condition gives
\begin{equation}
m=\frac{\chi^0_S}{1-I_S\chi^0_S}\,h\equiv\mathcal{A}_S\,\chi^0_S\,h .
\end{equation}
Iterating the self-consistency condition instead of solving it reproduces the diagrammatic content of the RPA,
\begin{equation}
\begin{split}
\chi_S&=\chi^0_S+\chi^0_S I_S\chi^0_S+\chi^0_S I_S\chi^0_S I_S\chi^0_S+\cdots\\
&=\sum_{n=0}^{\infty}\chi^0_S\bigl(I_S\chi^0_S\bigr)^n,
\end{split}
\end{equation}
the chain of electron--hole bubbles connected by the on-site interaction $I_S$ (Fig.~S2(c)); the series converges for $I_S\chi^0_S<1$ and diverges at the Stoner instability $I_S\chi^0_S=1$, where $\mathcal{A}_S\to\infty$.
For the orbital channel the same steps with $m\to m_L=n_{+1}-n_{-1}$ (the orbital polarization of the doublet $m=\pm1$ per spin direction), $I_S\to I_L$, and $\chi^0_S\to\chi^0_L$ give
\begin{equation}
\mathcal{A}_{L,S}=\bigl[1-I_{L,S}\chi^0_{L,S}\bigr]^{-1}.
\label{eq:SLS}
\end{equation}
The enhancement is thus fixed by the product $I\chi^0$ of the Stoner constant of Sec.~S4.2 and the bare susceptibility of the band; the factor $\tfrac12$ in $v\leftrightarrow I_S/2$ below reflects the normalization $E_{\rm pol}=-\tfrac{I}{4}m^2$ with $m=2S_z$.
For the spin channel the series is the electron--hole ladder of Izuyama, Kim, and Kubo \cite{IKK1963},
\begin{equation}\chi'_{-+}=\frac{\Gamma}{1-v\Gamma},\qquad v\leftrightarrow\tfrac{I_S}2,\end{equation}
whose static limit is the Stoner molecular field, whose zeros are spin waves, and which is cut off by the Stoner continuum; near $T_c$ it has the form
\begin{equation}1-vR(0,0)=A(T-T_c).\end{equation}
The orbital channel is isomorphic, with $\uparrow,\downarrow$ replaced by $m=\pm1$ and $v$ by $I_L/2$.
The spin paramagnon that enters the readout, Fig.~S2(c), is the chain of an odd number of bubbles between the $\uparrow$ and $\downarrow$ lines, which for a Hubbard-type on-site interaction is the series of Nakajima \cite{Nakajima1973} that enhances the triplet channel and suppresses the singlet one; its sum is
\begin{equation}\frac{x}{1-x^2}=\frac{x\,\mathcal{A}_S}{1+x},\qquad x=I_S\chi^0_S,\end{equation}
which near the instability, $x\to1$, reduces to $\mathcal{A}_S/2$; the enhancement factor used in all expressions is the RPA form $\mathcal{A}_S$ of Eq.~\eqref{eq:SLS}, and the odd chain is the part of it that enters the readout.
That the triplet channel is the one easily enhanced follows from the on-site interaction of Sec.~S4.1: the first Hund rule enters with four times the weight of the second, so the spin paramagnon reaches its instability before the orbital Stoner factor does.
The susceptibility also fixes where the enhancement is evaluated.
In the absorption step the factor at the phonon resonance $(q_*,\omega_*)$, $\mathcal{A}_L(q_*,\omega_*)$, raises the absorption rate of the rotation vertex and enters $\Pconv$, but drops out of the transfer per phonon; in the accumulation and readout steps the uniform static limits $\mathcal{A}_L(0,0)$ and $\mathcal{A}_S(0,0)$ amplify the stored polarization.
The extension of the same paramagnon to the particle--particle channel, the attraction between triplet pairs and the repulsion between singlet pairs, is the result of Ref.~\cite{Nakajima1973} and is not pursued here.

\section{Details on Step (6): SOC vertex and spin readout}

\subsection{Readout relations}
\label{sec:readout}
In the readout, SOC enters only as a single vertex $\lambda\hat{\boldsymbol L}\cdot\hat{\boldsymbol S}$.
Since the on-site interaction has no cross term (Sec.~S5), the orbital and the spin channel are enhanced independently, by $\mathcal{A}_L$ and $\mathcal{A}_S$ of Eq.~\eqref{eq:SLS}, and the vertex is dressed by $\mathcal{A}_L$ on its orbital side and by $\mathcal{A}_S$ on its spin side,
\begin{equation}
\mathcal{A}_S\langle S_z\rangle_0=\lambda\mathcal{A}_L\mathcal{A}_S\chi^0_S\langle L_z\rangle_0 ,\qquad\lambda_{\rm eff}=\lambda\mathcal{A}_L\mathcal{A}_S ,
\end{equation}
which is Eqs.~(5) and (6) of the main text; $\lambda$ itself is not renormalized.
For $\lambda=0$ no spin is induced: $\mathcal{A}_S$ survives but has no direction.

Packets injected at the rate $\nu$ deliver orbital angular momentum at the rate $\Phi_L=\nu\,Y\hbar$ (for a current density $j$ through a cross section $A$, $\nu/A=j/e$), of which the fraction $\eta=\lambda_{\rm eff}\chi^0_S$ is converted into spin, $P=2\eta Y$ and $\dot S_z^{\rm gen}=\nu(\hbar/2)P$ as in the main text.
Units: in the readout chain of the main text, Eqs.~(5)--(8), the angular momenta $\langle L_z\rangle_0$ and $\langle S_z\rangle_0$ carry their dimension, the SOC constant $\lambda$ is an energy, the effective field $h_{\rm eff}=\lambda\mathcal{A}_L\langle L_z\rangle_0/\hbar$ is an energy, and $\chi^0_S$ is the bare spin susceptibility in units of $\hbar$ per unit energy, $\langle S_z\rangle_0=\hbar\chi^0_Sh_{\rm eff}$; the angular-momentum operators of the multiplet algebra (Secs.~S4 and S5 and Appendix~C of the main text) are dimensionless, in units of $\hbar$.

In the steady state the spin relaxes with the spin relaxation time $\tau_s$, which gives the spin density
\begin{equation}
S_z^{\rm ss}=\eta\,\Phi_L\,\tau_s/V=\eta\,\nu\,Y\hbar\,\tau_s/V .
\label{eq:Sss}
\end{equation}
Here $V$ is the volume of the pumped region. The magnetization of the main text, Eq.~(8), is accordingly
\begin{equation}
M_z=-(\mu_B/\hbar)(\nu/V)\bigl[\tau_L\mathcal{A}_L\langle L_z\rangle_0+2\tau_s\mathcal{A}_S\langle S_z\rangle_0\bigr],
\end{equation}
where $\tau_L$ is the relaxation time of the orbital moment; in general $\tau_L\ll\tau_s$, since the orbital moment relaxes through the anisotropic potential of the neighboring ions, which quenches it.
Several consequences follow.
(i) $P$ does not depend on the injection rate $\nu$, whereas $\dot S_z^{\rm gen}$ and $S_z^{\rm ss}$ are linear in it.
(ii) $Y$ contains no SOC and, for a single resonance, not even the coupling constant (Sec.~S3.1); all the dependence on the readout site and on $\mathcal{A}_S$ is in $\eta$ (through $\lambda\mathcal{A}_S$, Sec.~S6.2).
(iii) $P\le1$ requires $2\eta Y\le1$.
(iv) $P$ is the yield of generated spin per injected packet; its relation to the polarization $(I_\uparrow-I_\downarrow)/(I_\uparrow+I_\downarrow)$ of a transmitted current depends on how the spin accumulated within $\tau_s$ is sampled by the outgoing electrons, and is not derived here.

\subsection{Readout site}
The coupling $\lambda$ is the atomic SOC that the orbitally polarized electrons meet first, not a Rashba coupling: inside the medium (for Te, the $5p$ shell, $0.3$--$0.5$~eV; for Nb, the $4d$ shell, $\sim0.1$~eV) or at an interface or electrode (Au, Pt, Ni; with Ti or Cu the angular momentum remains an orbital current).
In molecular CISS,
\begin{equation}
\langle S_z\rangle=\mathcal{A}_S^{({\rm det})}\chi_S^{0\,({\rm det})}\lambda^{({\rm det})}\,\mathcal{A}_L^{({\rm med})}\langle L_z\rangle_0^{({\rm med})},
\end{equation}
with ``med'' the medium and ``det'' the detector side; the electrode carries $\lambda$ and, for Ni, a spontaneous-magnetization $\mathcal{A}_S^{({\rm det})}$.
The accumulation scenario survives wherever the orbital accumulation reaches the readout point within the orbital diffusion length $\ell_L$ (extended by disorder \cite{Sohn2024}) and the readout point carries an atomic SOC.
The Hund coupling is the same in all materials; $\mathcal{A}_S$ can be strongly enhanced where the magnetic instability is strong, as in CrNb$_3$S$_6$ near $T_c$.

\section*{Part II: Temperature dependence and experimental signatures}

\section{Temperature dependence of the yield}
\label{sec:temp}
For a phonon that can split into two phonons of lower frequency the dominant lattice process is the Klemens decay \cite{Klemens1966,Balkanski1983},
\begin{equation}
\Gph(T)=\Gamma_0[1+2n(\hbar\omega_\mu/2)]=\Gamma_0\coth(\hbar\omega_\mu/4k_BT),
\end{equation}
linear in $T$ at room temperature.
This form is used for the phonon side at all temperatures, with a single $\Gamma_0$ and $\hbar\omega_\mu$.
The conversion probability is
\begin{equation}
\Pconv(T)=\Bigl[1+\frac{\Gamma_0}{\gel(T)}\coth\left(\frac{\hbar\omega_\mu}{4k_BT}\right)\Bigr]^{-1},
\label{eq:PT}
\end{equation}
and $Y\propto\Pconv$, because $|\al_\mu|^2$ is independent of temperature and, for a single resonance, so is $\langle\Delta L_z\rangle_\mu$ (Sec.~S3.1).

The electron side is the golden-rule absorption of the phonon $(\kappa_\mu,\omega_\mu)$ by electron--hole pairs. For a transition from an electronic state of energy $\varepsilon$ to one of energy $\varepsilon+\hbar\omega_\mu$, the thermal factor is
\begin{align}
f(\varepsilon)-f(\varepsilon+\hbar\omega_\mu)&=\bigl[1-e^{-\hbar\omega_\mu/k_BT}\bigr]\notag\\
&\quad\times f(\varepsilon)\bigl[1-f(\varepsilon+\hbar\omega_\mu)\bigr],
\label{eq:pauli}
\end{align}
where $f$ is the Fermi function. We call the energy interval $[\varepsilon,\varepsilon+\hbar\omega_\mu]$ the Pauli window of the transition.
If the window contains $E_F$ with margins much larger than $k_BT$, the factor is unity and the absorption does not depend on temperature (in a three-dimensional metal, where many windows contribute, the factor integrates to $\hbar\omega_\mu$ independently of $T$, Sec.~S3.1). This is the constant electron side, $\gel=\gamma_0$.
If the window lies entirely above or below $E_F$, and its nearest edge is at the energy distance $E_a\gg k_BT$ from $E_F$, the factor is $\simeq e^{-E_a/k_BT}$, and the electron side is activated, $\gel=\gamma_0e^{-E_a/k_BT}$.
Thus ``gapped'' refers only to the energy distance $E_a$ between the resonant transition and $E_F$. It does not mean that the material is an insulator; a metal has resonant strings of both kinds.

Two results used in the main text follow.
(i) For a constant electron side with $g\equiv\Gamma_0/\gamma_0$, the flat accumulation regime crosses over to $1/(1+bT)$ at $k_BT\gtrsim\hbar\omega_\mu/4$ with $b=4k_Bg/\hbar\omega_\mu$; the response limit $g\gg1$ gives
\begin{equation}
\Pconv\simeq g^{-1}\tanh(\hbar\omega_\mu/4k_BT).
\end{equation}

(ii) For a gapped electron side,
\begin{equation}
\Pconv=[1+g\,e^{E_a/k_BT}\coth(\hbar\omega_\mu/4k_BT)]^{-1};
\end{equation}
at $k_BT\gg\hbar\omega_\mu/4$ the denominator contains $T\,e^{E_a/k_BT}$, which has its minimum at $k_BT=E_a$, so $Y$ has a maximum at $T_{\rm peak}\simeq E_a/k_B$ for $E_a\gg\hbar\omega_\mu/4$ (numerically, $E_a=15$~meV and $\hbar\omega_\mu=10$~meV give $T_{\rm peak}=177$~K against $E_a/k_B=174$~K); the form $e^{-E_a/k_BT}/T$ is the limit $g\,e^{E_a/k_BT}\coth\gg1$.
A measured temperature dependence is described by one $\hbar\omega_\mu$ and $\Gamma_0$ on the phonon side and an electron side that is a sum of channels: a constant one, $\gamma_0$, for each open Pauli window, and an activated one, $\gamma_ke^{-E_k/k_BT}$, for each window at the energy distance $E_k$ from $E_F$; which channels are present is decided by the positions of the resonant windows relative to $E_F$.
Table~\ref{tab:fp} lists the characteristic signatures of the three types next to those of the Edelstein effect, which has no memory time; in the three types the memory time $\tau_\mu=1/(\gel+\Gph)$ is set by the sum of the two rates, whereas the type is set by their ratio, and the memory is observable when $\tau_\mu$ exceeds the time resolution of the measurement.

\begin{table*}[t]
\centering
\caption{Characteristic signatures of the three types of CAMP and of the Edelstein effect.}
\label{tab:fp}

\begin{tabular}{p{3.2cm}p{2.8cm}p{2.8cm}p{2.8cm}p{3.0cm}}
\toprule
 & accumulation & response & activated & Edelstein\\
\midrule
signal after the current is cut off & decays over $\tau_\mu$ & decays over $\tau_\mu$ & decays over $\tau_\mu$ & none\\
$T$ dependence of $Y$ & flat & $\propto1/T$ & $e^{-E_a/k_BT}/T$ & does not grow with $T$\\
$E_F$ sweep & peak of width $\sim k_BT$ & peak, height $\propto1/T$ & peak when the window opens & smooth\\
length dependence & orbital diffusion length $\ell_L$ & same & same & independent (bulk)\\
dependence on SOC of $Y$ & none & none & none & $\propto\lambda$\\
circular dichroism & enhanced by $\mathcal{A}_L\mathcal{A}_S$ & same & same & without enhancement\\
\bottomrule
\end{tabular}
\\[4pt]
\end{table*}

\section{Nonlocal signal of NbSi$_2$}
\label{sec:NbSi2}

We apply the rule of Sec.~S7 to the temperature dependence of the nonlocal signal measured at the center of a polycrystalline NbSi$_2$ sample.
The data are the 35 values of $Y(T)/Y(300\,{\rm K})$ read from Fig.~4(b) of Shishido \textit{et al.}~\cite{Shishido2021}, and we assume that the measured voltage ratio $V_{xy}/V_{xx}$ can be identified with the yield ratio $Y(T)/Y(300\,{\rm K})$.

NbSi$_2$ is a metal, so the Pauli window is open for at least one resonant string and the electron side contains a constant channel.
We allow one further channel whose window opens thermally at an energy $E_a$, so that
\begin{equation}
\gel(T)=\gamma_0\bigl[1+r\,e^{-E_a/k_BT}\bigr],\qquad r\equiv\frac{\gamma_1}{\gamma_0},
\label{eq:two}
\end{equation}
and the phonon side is the Klemens form of Eq.~\eqref{eq:PT} with one $\hbar\omega_\mu$. With $g_0\equiv\Gamma_0/\gamma_0$,
\begin{align}
\frac{Y(T)}{Y(300\,{\rm K})}&=\frac{P_{\rm conv}(T)}{P_{\rm conv}(300\,{\rm K})},\notag\\
P_{\rm conv}(T)&=\Bigl[1+\frac{g_0}{1+r\,e^{-E_a/k_BT}}\coth\frac{\hbar\omega_\mu}{4k_BT}\Bigr]^{-1}.
\label{eq:low}
\end{align}
We take $\hbar\omega_\mu=20$~meV, so that $\hbar\omega_\mu/4k_B\simeq58$~K (a fit parameter, not an independently identified resonant phonon).
The model then has three parameters, $g_0$, $r$, and $E_a$, for 35 data points.

\begin{figure}[tb]
\centering
\includegraphics[width=\columnwidth]{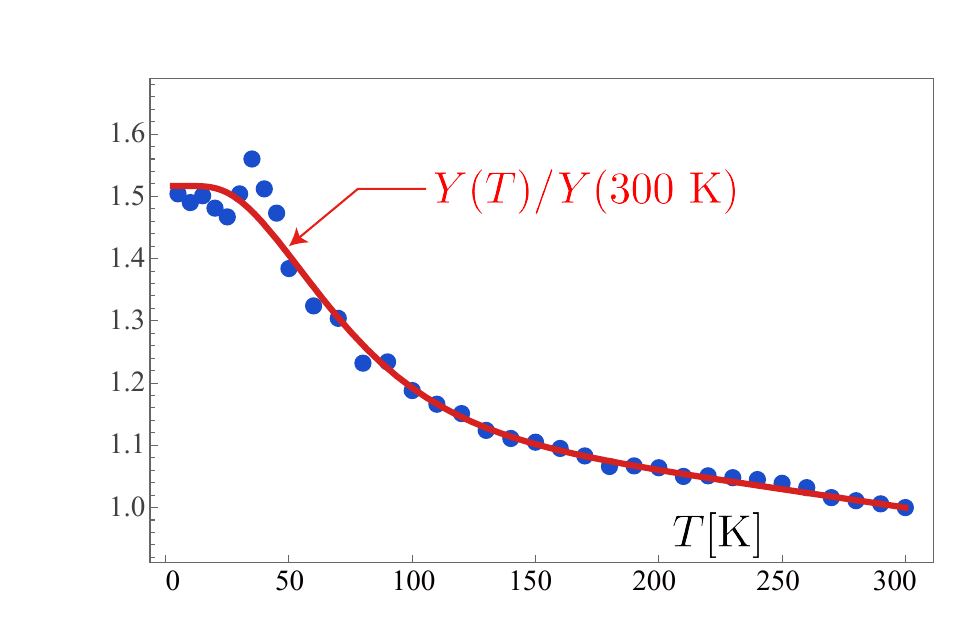}
\caption{Temperature dependence of the nonlocal signal at the center of a polycrystalline NbSi$_2$ sample.
The 35 values of $Y(T)/Y(300\,{\rm K})$ read from Fig.~4(b) of Ref.~\cite{Shishido2021} (blue dots) and Eq.~\eqref{eq:low} (red) with the electron side of Eq.~\eqref{eq:two} (see Secs.~S7 and S8), $\hbar\omega_\mu=20$~meV, $g_0=0.58$, $r=4.1$, and $E_a=32$~meV.}
\label{fig:S4}
\end{figure}

Fitting all 35 points with $g_0$, $r$, and $E_a$ free gives
\begin{equation}
\begin{gathered}
g_0=0.58\pm0.09,\qquad r=4.1\pm0.2,\\
E_a=32\pm2~\text{meV},
\end{gathered}
\end{equation}
with an rms deviation of $0.023$ (Fig.~\ref{fig:S4}).
A constant electron side alone ($r=0$) cannot follow the data: the rms deviation is $0.069$ even when $\hbar\omega_\mu$ is also varied, and a linear increase $\gel\propto1+T/T_1$ gives $0.056$. A second, thermally activated channel is required.

At low temperature the lattice loss is frozen at its zero-point value, $\Gph\to\Gamma_0$, and $\Pconv$ is constant, $\Pconv(5\,{\rm K})=0.63$: the signal is flat up to $T\sim\hbar\omega_\mu/4k_B$.
Above it $\Gph$ grows linearly with $T$ and $\Pconv$ decreases, to $\Pconv(300\,{\rm K})=0.42$.
The activated channel raises $\gel$ only slowly, relative to its value at $5$~K by $1.0$ at $60$~K, $1.1$ at $100$~K, $1.6$ at $200$~K, and $2.2$ at $300$~K; it flattens the high-temperature part of the decrease relative to the linear form $1/\Pconv\simeq1+bT$, $b=4k_Bg/\hbar\omega_\mu$, of a constant electron side.

The fit determines the activated channel, but $\hbar\omega_\mu$ and $g_0$ are degenerate.
Refitting $g_0$, $r$, and $E_a$ at fixed $\hbar\omega_\mu$ gives rms deviations of $0.031$, $0.028$, $0.025$, $0.023$, $0.021$, $0.021$ for $\hbar\omega_\mu=8$, $12$, $16$, $20$, $25$, $30$~meV, and $0.038$ and $0.077$ for $40$ and $60$~meV.
Over $16$--$30$~meV, $g_0$ rises from $0.31$ through $0.58$ ($20$~meV) to $\gtrsim40$ ($30$~meV), while $r=3.8$--$4.6$ and $E_a=31$--$34$~meV change little.
The data therefore fix the activated channel but not whether the plateau is of the accumulation ($g_0\ll1$) or the response ($g_0\gg1$) type.

The form of the electron side is motivated as follows. A metal has many resonant strings, each with its own window (Sec.~S7), so that in general
\begin{equation}
\gel(T)=\sum_s\gamma_sF_s(T),
\end{equation}
with $F_s\simeq1$ for a window that contains $E_F$ and $F_s\simeq e^{-E_s/k_BT}$ for a window at the energy distance $E_s$ from $E_F$.
The two channels of Eq.~\eqref{eq:two} are the two dominant classes of this sum, a window at $E_F$ and a window at a single distance $E_a$; 35 data points with a scatter of a few percent do not resolve more.
Here $E_a$ is not a gap of the electronic spectrum. NbSi$_2$ is a metal, and $E_a$ is the energy distance of a resonant transition from $E_F$.
The temperature dependence therefore provides an experimental measure of that distance. In the present fit, $E_a=32\pm2$~meV ($\simeq370$~K) places the transition of the second channel about $32$~meV from $E_F$, on a side that the temperature dependence alone does not determine.
The value can be compared with the band structure and with angle-resolved photoemission, and, since a shift of $E_F$ changes $E_a$, it can be tested by an $E_F$ sweep (Table~\ref{tab:fp}).

The model does not reproduce the sharpness of the knee at $35$--$60$~K: the residuals (data minus fit) are $+0.08$, $+0.06$, $+0.04$ at $35$, $40$, $45$~K and $-0.02$, $-0.03$ at $50$, $60$~K, whereas for $T\ge100$~K they stay within $0.01$.

The comparison is qualitative support for the scheme: the data show the flat plateau and the decrease expected from Eq.~\eqref{eq:PT}. This is the expected structure: $|\al_\mu|^2$ and, for a single resonance, $\langle\Delta L_z\rangle_\mu$ carry no temperature dependence (Sec.~S7), and the readout factor $\eta$ of a diamagnetic metal is temperature independent away from a magnetic instability, so the measured temperature dependence is that of $\Pconv$ alone and probes the reservoir, not the readout.

\section*{Part III: Extensions}

\section{Lower symmetry, polarization, and circular dichroism}
This section treats the pump in structures of lower symmetry (Sec.~S9.1), and the parallel polarization and magnetization together with the circular dichroism (Sec.~S9.2).

\subsection{Lower symmetry}
The one-electron pump
\begin{equation}
J_1=\sum_\mu\ell_{\mu,\parallel}|\al_\mu|^2,
\end{equation}
with $\ell_{\mu,\parallel}$ the axial component of $\ell_\mu$, is defined for any chiral structure with normal modes, periodic or finite, without a screw eigenvalue or a CAM label, and for a periodic structure the phase matching with the Bloch wave vector is as sharp for $P2$ or $222$ as for $P3_121$; this CAM-free form is the one used when the screw axis has no handedness of its own and the chirality resides in the atoms at general positions, as in CrNb$_3$S$_6$ (Sec.~S1.1).

The pump also distinguishes chiral from merely gyrotropic structures. In general $J_{1,i}=g_{ij}v_j$, with $g_{ij}$ an axial--polar tensor of the same form as the gyration tensor. In the enantiomorphic point groups $g_{ij}$ has a trace, the pseudoscalar, so that the angular momentum has a component along the injection for every direction of $\boldsymbol v$, and this longitudinal component is what the sign rule of the main text describes. In a gyrotropic but achiral group such as $C_{2v}$ the mirror planes remove the trace and leave only off-diagonal components ($g_{xy}$, $g_{yx}$): the longitudinal angular momentum vanishes, and injection along $x$ produces an angular momentum along $y$, whose sign reverses when the injection direction is reflected in either mirror plane and which vanishes for injection along a mirror plane or along the twofold axis. A CISS-type signal in such a structure is therefore expected to depend sensitively on the direction of injection relative to the mirror planes, and a transverse rather than longitudinal spin is the signature that separates the gyrotropic from the chiral case.

\subsection{Parallel polarization and magnetization, and circular dichroism}
In a metal the static $\boldsymbol P$ is screened and the surviving polar vector is the current density $\boldsymbol j$, so that $\boldsymbol M\parallel\boldsymbol j$ is shared by CAMP and by the Edelstein effect.
In an insulating molecule or junction a charge displacement along the molecule is a transient current that kicks the ions and radiates chiral vibrations; their angular momentum is absorbed by the electron--hole excitations of the molecule, and the magnetization parallel to the polarization lags it by the vibrational memory time (there is no Klemens continuum in a molecule).
Since the orbital and spin polarizations and their fluctuations are enhanced by $\mathcal{A}_L$ and $\mathcal{A}_S$, the circular dichroism, which probes them, is expected to be enhanced as well; summing and subtracting the signals of the two enantiomers separates the enantiomer-odd from the enantiomer-even components.
This is not evaluated here.

\section*{Part IV: Material map}

\section{Material map}
The materials of the main text are classified by four properties, each of which enters a different step of the scheme.
(i) The shell of the electrons that carry the rotation vertex ($p$ or $d$) fixes the bright CAM sectors: $m=\pm1$ for the $p$ shell, $|m|\le2$ for the $d$ shell (Sec.~S1.1).
(ii) The helix type ($n_p$) fixes the handedness and the basis in which $J_1$ is written; for a screw axis with $p=n/2$ the handedness resides in the atoms at general positions and $J_1$ is written in the CAM-free form (Secs.~S1.1 and S9.1).
(iii) Whether the Pauli window of the resonant phonon is open, that is, whether the string of the band structure that satisfies $\varepsilon(k+q)-\varepsilon(k)=\hbar\omega_*$ reaches the Fermi level (or, in a molecule, the HOMO--LUMO transition), decides the type of Sec.~S7 through $\gel/\Gph$.
(iv) The magnetic state of the medium fixes $\mathcal{A}_S$ (Sec.~S5).
The classification is a framework, not a set of predictions: the resonant modes of the $d$-electron metals have not been identified, and the class of each material has to be decided by first-principles calculations of the phonon spectrum, the resonant strings, and $\gel/\Gph$.

\begin{table*}[t]
\centering
\caption{Materials discussed in the main text and their structural parameters.}

\begin{tabular}{p{3.3cm}p{2.8cm}p{1.0cm}p{6.0cm}}
\toprule
material & helix & shell & remark\\
\midrule
Te & $3_1$ ($3_2$) & $p$ & $m=\pm1$ bright\\
NbSi$_2$, TaSi$_2$ & $6_2$ ($6_4$) & $d$ & $|m|\le2$ bright (Sec.~S1.1)\\
CrNb$_3$S$_6$ & $6_3$ & $d$ & $p=n/2$: handedness in the twisted S$_6$ octahedra at general positions; CAM-free form (Sec.~S9.1)\\
$\alpha$ helix, DNA & $18_5$; $n\simeq10.5$ & $p$ & large-$n$ limit\\
\bottomrule
\end{tabular}
\end{table*}

Molecular helices are the limit of large $n$: the $\alpha$ helix is an $18_5$ screw (eighteen residues in five turns, $3.6$ residues per turn \cite{Pauling1951,Cochran1952}), and in B-DNA the $10.5$ base pairs per turn are not a ratio of small integers, so that no short repeat exists and effectively $n\to\infty$.
In this limit $\mathbb Z_n\to\mathbb Z$: the CAM sectors become dense, $\kappa=q+2\pi m/(n\tau)$ fills the extended zone, and the CAM becomes the angular momentum of a continuous helix.
The selection rule of the rotation vertex is unchanged, $m\to m\pm1$, so out of this quasi-continuum only $|m|\le l_{\rm shell}$ is read out, $|m|\le1$ for $p$ orbitals (Sec.~S1.1); at the same time the number of Umklapp orders $j$ grows with $n$, and the phase matching becomes easy to satisfy.

The electron side of a molecule differs from that of a metal in two respects, and both enter the temperature dependence.
First, the electronic spectrum is discrete.
The Pauli window of Sec.~S3.1 is open only if the injected electron finds, within its own manifold of levels, a partner level separated from it by $\hbar\omega_\mu$; in general it does not, and the phonon is absorbed through thermally populated vibronic or multiplet levels that bring a pair into resonance.
This is the gapped electron side of Sec.~S7, $\gel=\gamma_0e^{-E_a/k_BT}$, where $E_a$ is the energy mismatch between the level spacing of the injected electron and $\hbar\omega_\mu$; it is not the HOMO--LUMO gap of the neutral molecule, which would suppress the channel entirely at room temperature.
The yield is then of the activated type (Sec.~S7): it rises as $e^{-E_a/k_BT}$ and has its maximum at $k_BT\simeq E_a$.
Second, a molecule has no phonon continuum, so the loss $\Gph$ is the rate of intramolecular vibrational relaxation rather than the Klemens form; the rule of a common $\hbar\omega_\mu$ on the phonon side does not apply.

A molecule has no Fermi sea, so the Stoner denominator $1-I\chi^0$ is replaced by the Hund-multiplet splitting of the excited configuration, and the enhancement is governed by the fluctuations of the spin and orbital polarization of that configuration (Sec.~S4.2); a large $\lambda_{\rm eff}$, with $\lambda$ itself unchanged, may therefore be available in molecules as well.

\end{document}